\documentclass[useAMS,usenatbib]{mn2e}

\usepackage{color}
\usepackage{colortbl}
\usepackage{multirow}
\usepackage{dcolumn}
\usepackage{amsmath}
\usepackage{amssymb}
\usepackage{graphicx}
\usepackage{epsfig}
\usepackage{changebar}
\usepackage{natbib}
\usepackage{multirow}
\usepackage{subfigure}
\usepackage{txfonts}
\usepackage{rotating}

\usepackage{longtable,lscape}

\newcolumntype{d}[1]{D{.}{\cdot}{#1}}

\newcolumntype{.}{D{.}{.}{-1}}

\newcommand{\lsun}{L$_\odot$}
\newcommand{\msun}{M$_\odot$}

\newcommand{\vlsr}{V$_{\rm{LSR}}$}

\newcommand{\kms}{km\,s$^{-1}$}

\newcommand{\Tk}{$T_{\mathrm{k}}$}

\newcommand{\nhthree}{NH$_3$}

\title[NH$_3$ and H$_2$O maser analysis of massive star forming regions]{The RMS Survey: Ammonia and water maser analysis of massive star forming regions.\thanks{Tables\,2 and 5 and full versions of Figs.\,3 and 7 are only available in electronic form at the CDS via anonymous ftp to cdsarc.u-strasbg.fr (130.79.125.5) or via http://cdsweb.u-strasbg.fr/cgi-bin/qcat?J/A+A/.}}
\author[J. S. Urquhart et al.]{
J.\,S.\,Urquhart$^{1}$\thanks{E-mail:
James.Urquhart@csiro.au (CASS)}, L.\,K.\,Morgan$^{2}$, C.\,C.\,Figura$^{3}$, T.\,J.\,T.\,Moore$^{2}$, S.\,L.\,Lumsden$^{4}$, M.\,G.\,Hoare$^{4}$, \newauthor  R.\,D.\,Oudmaijer$^{4}$, J.\,C.\,Mottram$^{5}$, B.\,Davies$^{6}$, M.\,K.\,Dunham$^{7}$ \\
$^{1}$
CSIRO Astronomy and Space Science, P.O. Box 76, Epping, NSW 1710, Australia \\
$^{2}$Astrophysics Research Institute, Liverpool John Moores University, Twelve Quays House, Egerton Wharf, Birkenhead, CH41\,1LD, UK\\ 
$^{3}$Wartburg College, 100 Wartburg Blvd, Waverly, IA 50677, USA \\
$^{4}$School of Physics and Astrophysics, University of Leeds, Leeds, LS2\,9JT, UK \\
$^{5}$School of Physics, University of Exeter, Exeter, EX7 4QL, UK\\
$^{6}$Institute of Astronomy, University of Cambridge, Madingley Road, Cambridge, CB3\,0HA, UK  \\
$^{7}$Department of Astronomy, Yale University, P.O. Box 208101, New Haven, CT 06520-8101\\
}

\begin{document}

\date{Accepted ??. Received ??; in original form ??}

\pagerange{\pageref{firstpage}--\pageref{lastpage}} \pubyear{2009}

\maketitle

\label{firstpage}

\begin{abstract}
The Red MSX Source (RMS) survey has identified a sample of $\sim$1200 massive young stellar objects (MYSOs), compact and ultra compact H{\sc ii} regions from a sample of $\sim$2000 MSX and 2MASS colour selected sources. We have used the 100-m Green Bank telescope to search for 22-24\,GHz water maser and ammonia (1,1), (2,2) and (3,3) emission towards $\sim$600 RMS sources located within the northern Galactic plane. We have identified 308 H$_2$O masers which corresponds to an overall detection rate of $\sim$50\,per\,cent. We find no significant difference in the detection rate for H{\sc ii} regions and MYSOs which would suggest that the conditions required to produce maser emission are equally likely in both phases. Comparing the detection rates as a function of luminosity we find the H$_2$O detection rate has a positive dependence on the source luminosity, with the detection rate increasing with increasing luminosity. 

We detect ammonia emission towards 479 of these massive young stars, which corresponds to $\sim$80\,per\,cent. Ammonia is an excellent probe of high density gas allowing us to measure key parameters such as gas temperatures, opacities, and column densities, as well as providing an insight into the gas kinematics. The average kinetic temperature, FWHM line width and total NH$_3$ column density for the sample are approximately 22\,K, 2\,\kms\ and $2\times 10^{15}$\,cm$^{-2}$, respectively. We find that the NH$_3$ (1,1) line width and kinetic temperature are correlated with luminosity and finding no underlying dependence of these parameters on the evolutionary phase of the embedded sources, we conclude that the observed trends in the derived parameters are more likely to be due to the energy output of the central source and/or the line width-clump mass relationship. 

The velocities of the peak H$_2$O masers and the NH$_3$ emission are in excellent agreement with each other, which would strongly suggest an association between the dense gas and the maser emission. Moreover, we find the bolometric luminosity of the embedded source and the isotropic luminosity of the H$_2$O maser are also correlated. We conclude from the correlations of the cloud and water maser velocities and the bolometric and maser luminosity that there is a strong dynamical relationship between the embedded young massive star and the H$_2$O maser.

\end{abstract}

\section{Introduction}

Massive young stellar objects (MYSOs) and ultra compact (UC) H{\sc ii} regions are two of the earliest phases in the lives of OB stars. The MYSO phase begins when heating of the proto stellar envelope increases visibility of the object in the mid-infrared, and ends once the central star begins to ionize its surrounding environment and forms an UC\,H{\sc ii} region. These two phases are physically distinct from the earlier hot molecular core phase, which is not generally detectable at mid-infrared wavelengths (e.g., \citealt{de-buizer2002}). MYSOs also possess strong ionized stellar winds (e.g., \citealt{bunn1995}); however, the radio emission from these winds is relatively weak ($\sim$1\,mJy at 1\,kpc; \citealt{hoare1994}) and easily distinguishable from the slightly later radio-loud UC\,H{\sc ii} region phase. It is likely that many MYSOs and UC\,H{\sc ii} regions are still accreting, as evidenced by their almost ubiquitous association with powerful bipolar outflows (\citealt{lada1985}).

Models suggest that high levels of accretion in the early stages of a massive star's development lead to a `swelling up' of the protostellar core due to trapped entropy (e.g., \citealt{yorke2008,hosokawa2009,hosokawa2010}). The length of this swollen phase is comparable to the Kelvin-Helmholz timescale, which for stars with a current mass of $\ga$20\,\msun\ is only a few thousand years. This means that once the accreted mass exceeds $\sim$20\,\msun\ the star arrives on the main-sequence, even if accretion is still ongoing \citep[see also][]{mckee2003}. At this point, the star begins to ionize its surroundings and becomes a UC\,H{\sc ii} region. This means that the MYSO phase is likely to be very brief ($\sim 10^{5}$\,yrs) and is limited to objects with {\it current} masses lower than $\sim$20\,\msun\ (\citealt{davies2011}). The UC\,H{\sc ii} region phase lasts a factor of 2-5 times longer, with a small dependence on the {\it final} stellar mass (\citealt{davies2011,mottram2011b}).

The Red MSX Source (RMS; \citealt{hoare2005}; \citealt{mottram2006}; \citealt{urquhart2007c}) Survey has established a large ($\sim$1200)  and well-selected sample of MYSOs and compact and UC\,H{\sc ii} regions, and a database of complementary multi-wavelength data.\footnote{www.ast.leeds.ac.uk/cgi-bin/RMS/RMS\_DATABASE.cgi.} We have used arcsecond-resolution mid-infrared imaging from the Spitzer GLIMPSE survey (\citealt{benjamin2003}) or our own ground-based imaging (e.g., \citealt{mottram2007}) to reveal multiple and/or extended sources within the MSX beam, as well as MYSOs in close proximity to existing H{\sc ii} regions. We have obtained arcsecond-resolution radio continuum with ATCA and the VLA (\citealt{urquhart_radio_south,urquhart_radio_north}) to identify UC\,H{\sc ii} regions and PNe, whilst observations of $^{13}$CO transitions (\citealt{urquhart_13co_south,urquhart_13co_north}) deliver kinematic distances and luminosities, which allow us to distinguish between nearby low- and intermediate-mass YSOs and genuine MYSOs. Finally we have obtained near-infrared spectroscopy (e.g., \citealt{clarke2006}) which allows us to distinguish the more evolved stars.

With the source classification effectively complete, the next step is to examine the global characteristics of this Galaxy-wide sample of massive young stars. This involves determining the physical and chemical nature of the environment as a way of gauging the evolutionary status of our sample of MYSOs and H{\sc ii} regions. In this paper we present the results of a set of thermal ammonia (NH$_3$) and water (H$_2$O) maser observations made with the 100-m Green Bank telescope (GBT) towards a sample of $\sim$600 young massive stars.

Ammonia is an excellent tracer of high-density gas ($\sim$10$^4$\,cm$^{-3}$; \citealt{evans1999,stahler2005}), and is relatively unaffected by depletion at lower temperatures compared to other common molecular tracers such as CO (\citealt{bergin1997}). The NH$_3$ (1,1), (2,2) and (3,3) inversion transitions are normally collisionally excited and the rotational temperature of the gas can be determined from the intensity ratio of any two inversion transitions. The inversion transition splits into 18 separate hyperfine components (\citealt{ho1983}), usually resolved into five distinct components, the ratio of which can be used to calculate the optical depth of the transition.

H$_2$O masers are known to occur in both high and low-mass star-forming regions (e.g., \citealt{forster1999,claussen1996}) and are thus an important signpost of ongoing star formation. Water masers are generally thought to be associated with molecular outflows (\citealt{codella2004} and references therein). We will use the H$_2$O detection rates for our sample of MYSOs and H{\sc ii} regions to investigate their statistical association as a function of evolutionary phase and luminosity.

The structure of the paper is as follows: in Sect.\,2 we discuss the source selection, the observational setup and the data reduction procedure. In Sect.\,3 we use the reduced spectra to determine physical properties of the star-forming environments. In Sect.\,4 we present the results and discuss their implications. We present a summary of our results and highlight our main findings in Sect.\,5.

\section{Observations and data reduction}

\subsection{Source selection}

The multi-wavelength data sets compiled as part of the RMS survey have been used to classify the initial sample into a number of different source types (see \citealt{urquhart2007c} for an overview of the classification scheme). For these observations we have selected all RMS sources classified as a young stellar object or H{\sc ii} region. In some cases there is more than one embedded source located within the MSX beam ($\sim$18\arcsec) or a source can display traits of both H{\sc ii} regions and YSOs and is possibly in a transitional stage between the two; these sources are classified as `H{\sc ii}/YSO' and were also included in our sample. Finally, we have been unable to definitively classify approximately 10\,per\,cent of sources where the available data are contradictory; these are designated `young/old' to indicate the uncertainty associated with them. Although the majority of these sources are likely to be evolved stars, we include them to avoid exclusion of any genuine YSOs that may still be lurking in this group. 

This selection process produced a sample of 586 sources located in the northern Galactic plane (i.e., $10\degr < l < 180\degr$). This sample provides a complete census of compact and UC\,H{\sc ii} regions, and MYSOs in the first and second quadrants. We complement this representative sample with a further eleven sources located in the outer part of the southern Galactic plane (i.e., $220\degr < l < 240\degr$).

\subsection{GBT Observations}

\begin{table}

\begin{center}\caption{Observed transition frequencies and excitation temperatures.}
\label{tbl:line_frequencies}
\begin{minipage}{\linewidth}
\begin{tabular}{lcclc}
\hline \hline
 &  	\multicolumn{1}{c}{Frequency$^a$}  &\multicolumn{1}{c}{Energy$^b$} & \multicolumn{1}{c}{Maser$^c$}& Sensitivity\\
Transition & (GHz) & (K) & \multicolumn{1}{c}{or Thermal} & \\
\hline
NH$_3$ (1,1) 	& 23.694471  	& 23.4	& thermal & 52\,mK\\
NH$_3$ (2,2) 	& 23.722634  	& 64.9	& thermal & 50\,mK\\
NH$_3$ (3,3) 	& 23.870130  	& 124.5	& both    & 54\,mK\\
H$_2$O (6$_{1,6}-5_{2,3}$) 	& 22.235180  	& $\cdots$	& maser & 0.12\,Jy\\

\hline\\
\end{tabular}\\
$^a$ Transition rest frequencies have been taken from the Lovas catalogue available from http://physics.nist.gov/cgi-bin/micro/table5/start.pl.\\
$^b$ Energy above the ground state in equivalent temperatures taken from \citet{ho1983}.\\
$^c$ Indicates whether a transition is a thermal line, a maser or possibly both.
\end{minipage}

\end{center}
\end{table}

\begin{figure}
\begin{center}
\includegraphics[width=0.23\textwidth, trim= 30 0 10 20]{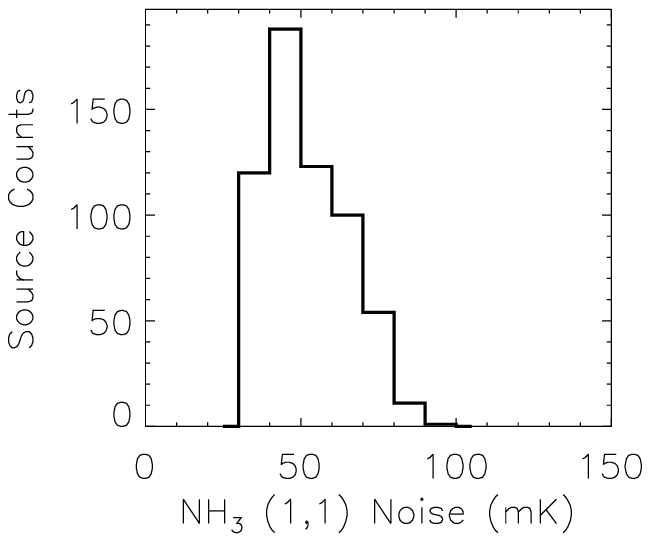}
\includegraphics[width=0.23\textwidth, trim= 30 0 10 20]{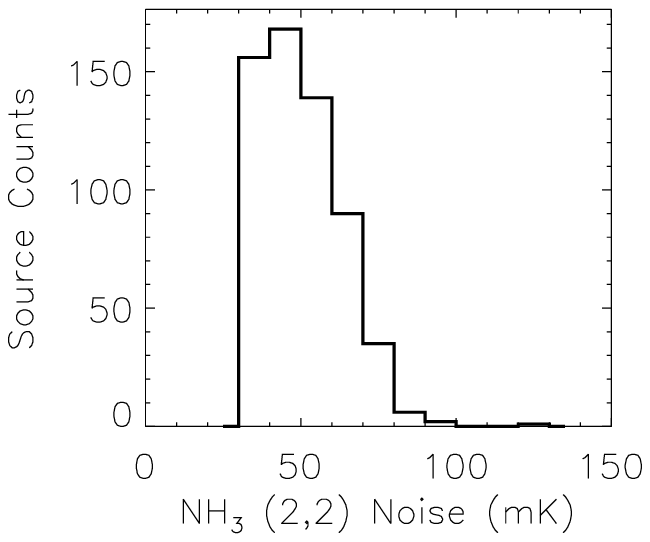}
\includegraphics[width=0.23\textwidth, trim= 30 0 10 20]{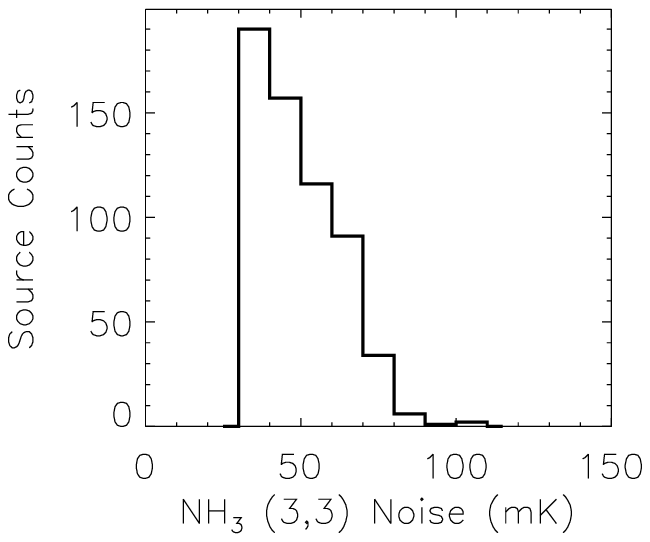}
\includegraphics[width=0.23\textwidth, trim= 30 0 10 20]{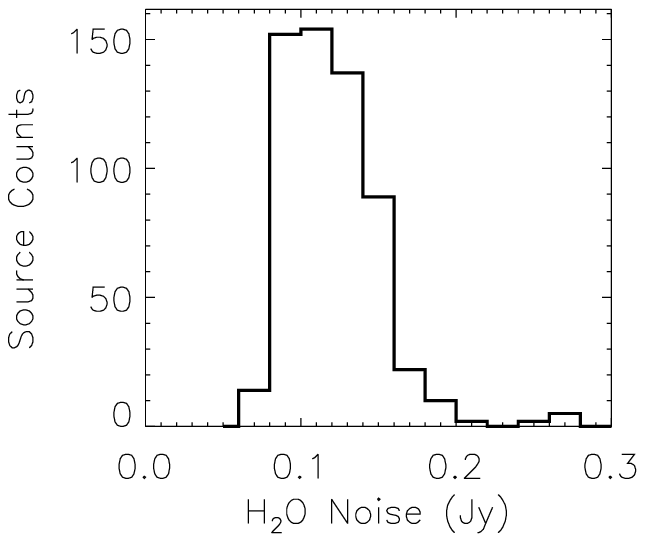}

\caption{\label{fig:noise} Plots of the noise distribution of the four observed lines.} 

\end{center}
\end{figure}

Observations were made of the NH$_3$ (1,1), (2,2) and (3,3) inversion transitions and the H$_{2}$O maser transition towards a sample of 597 young massive stars identified by the RMS survey (see Table\,\ref{tbl:line_frequencies} for transition rest frequencies). These observations were made during six sessions starting on the 25$^{\rm{th}}$ of November 2009 and ending on the 10$^{\rm{th}}$ of December 2010 using the Green Bank telescope (GBT), operated by the National Radio Astronomy Observatory\footnote{The National Radio Astronomy Observatory is a facility of the National Science Foundation operated under cooperative agreement by Associated Universities, Inc.}.

All three NH$_3$ transitions were observed simultaneously, eliminating many of the sources of uncertainty while allowing many key parameters such as excitation, rotation and kinetic temperatures, column densities, and optical depths to be calculated for a large sample of MYSOs and H{\sc ii} regions. The observations were performed in frequency-switch mode in order to remove sky contributions and a noise diode was observed in switching mode throughout the observations in order to achieve absolute flux calibration on the $T_{\rm{A}}^*$ scale to an accuracy of $\sim$10\,per\,cent.  The natural spectral resolution  of the raw data was 6.1\,kHz ($\sim$0.08\,\kms). A standard integration time of 70 seconds was used for each source, resulting in a typical r.m.s. noise of 100\,mK per channel with median system temperatures of 61\,K. Weather conditions were stable for all observing sessions: typical pointing offsets were $\sim$5\arcsec\ in azimuth and $\sim$3\arcsec\ in elevation with a half-power beam-width of $\sim$30\arcsec.

\subsection{Data reduction}
\label{sect:data_reduction}

The data were reduced using the GBTIDL\footnote{http://gbtidl.nrao.edu/} data analysis package. Bad scans were removed and channels outside the region of interest were discarded. Temperature scale corrections for atmospheric opacity were made using the zenith values provided from local weather models. A high-order polynomial was fitted to emission free channels and subtracted from all spectra to remove baseline anomalies before any emission features were fitted.   

The reduced spectra were Hanning-smoothed to a resolution of $\sim$0.32\,\kms\ and a sensitivity of 50\,mK\,channel$^{-1}$. In Fig.\,\ref{fig:noise} we present histograms of the noise distributions of the four observed transitions. Noise values are given in mK for the ammonia transitions; however, for the water masers we have converted the temperature scale to the more commonly used Jansky scale using the GBT gain value of 1.5\,K\,Jy$^{-1}$.\footnote{\label{gbtfootnote}The gain factor and main beam efficiency have been taken from `The Proposer's Guide for the Green Bank Telescope'.}   

The final reduced spectra were calibrated to the corrected antenna temperature scale ($T_{\rm{A}}^*$), and converted to the telescope-independent main-beam temperature scale ($T_{\rm{mb}}$), assuming a main-beam efficiency ($\eta_{\rm{mb}}$) of 0.89$^{4}$.

\begin{figure}
\begin{center}
\includegraphics[width=0.49\textwidth, trim= 0 0 0 0]{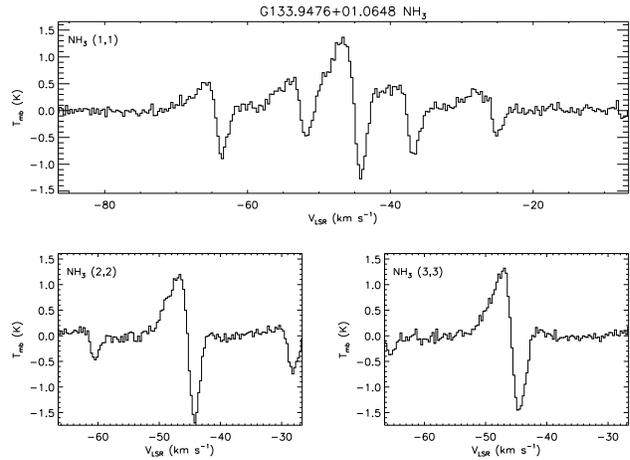}

\caption{\label{fig:nh3_complex} Ammonia spectra towards the MSX source G133.9476+01.0648 as an example of a source with both emission and absorption. The full version of this figure can be found in Fig\,\ref{fig:nh3_spectra_complex}.} 

\end{center}
\end{figure}

The NH$_3$ emission seen towards the vast majority of sources can be attributed to a single molecular cloud located along the line of sight.  Multiple clouds, identified through multiple spectral components, are seen towards only nine sources. More complex emission structure is seen towards six  where the spectra appear to be combinations of emission and absorption; these spectra indicate a bright continuum source in the beam (i.e., H{\sc ii} region). In all of these cases the velocity ranges of the absorption and emission overlap, making it difficult to obtain reliable parameters. We present the emission detected towards one of these sources in Fig.\,\ref{fig:nh3_complex} (the spectra seen towards all six of these sources can be found in Fig.\,\ref{fig:nh3_spectra_complex} of the Appendix). Since the results obtained by fitting these spectra are unlikely to be reliable, we simply present these spectra without further analysis. 

\section{Deriving physical properties}

In this section we will describe the methods used to determine physical properties of the dense environments in which young massive stars are forming.

\begin{landscape}

\begin{table}
\centering
\caption{Observed parameters.}
\label{tbl:nh3_parameters}
\begin{minipage}{\linewidth}
\scriptsize

\begin{tabular}{lcccccccccccccccccccc}
& &	  		& &&&& 		& \multicolumn{4}{c}{NH$_3$ (1,1)}& & \multicolumn{3}{c}{NH$_3$ (2,2)} && \multicolumn{3}{c}{NH$_3$ (3,3)} &  \\ 
\cline{9-12} \cline{14-16}\cline{18-20}
MSX Name$^{a}$&	\multicolumn{1}{c}{Type$^{b}$}	& 				 	 	RA  		&Dec &$T_{\rm{ex}}$ &$T_{\rm{rot}}$ &$T_{\rm{kin}}$ &Log[$N({\rm{NH3}})$] &T$_{\rm{mb}}$$^{c}$ & \vlsr & $\Delta$V & $\tau_{\rm{(m,1,1)}}$ &&T$_{\rm{mb}}$ & \vlsr & $\Delta$V && T$_{\rm{mb}}$ & \vlsr & $\Delta$V & H$_2$O\\
&	&(J2000) 	& (J2000) 	&  (K) &  (K) & (K) &  (cm$^{-2}$) &(K)  &(\kms) & (\kms) &          & &     (K) & (\kms) &(\kms) & &(K) & (\kms) & (\kms) & \\

\hline
G010.3844+02.2128	&	1	&	18:00:22.68	&	$-$18:52:08.0	&	4.5	&	14.8	&	16.0	&	15.0	&	1.2	&	5.54	&	0.72	&	1.11	& &	0.4	&	5.44	&	0.77	& &	$\cdots$	&	$\cdots$	&	$\cdots$	&	y	\\
G010.4413+00.0101	&	2	&	18:08:38.23	&	$-$19:53:57.4	&	3.7	&	16.8	&	18.5	&	15.7	&	0.6	&	66.37	&	2.63	&	1.54	& &	0.3	&	66.88	&	3.08	& &	0.2	&	66.28	&	4.54	&	n	\\
G010.4718+00.0206	&	2	&	18:08:39.67	&	$-$19:52:03.0	&	4.1	&	20.7	&	24.0	&	16.2	&	1.0	&	67.40	&	5.73	&	2.17	& &	0.8	&	67.55	&	5.90	& &	0.8	&	67.50	&	7.88	&	y	\\
G010.5067+02.2285	&	1	&	18:00:34.58	&	$-$18:45:17.6	&	5.7	&	14.5	&	15.5	&	15.4	&	2.4	&	21.79	&	1.31	&	1.83	& &	0.9	&	21.86	&	1.69	& &	0.2	&	21.94	&	3.24	&	y	\\
G010.6291$-$00.3385	&	2	&	18:10:19.32	&	$-$19:54:12.9	&	6.9	&	17.2	&	19.1	&	15.9	&	3.6	&	$-$4.54	&	1.86	&	3.39	& &	2.5	&	$-$4.51	&	2.57	& &	1.1	&	$-$4.42	&	3.19	&	n	\\
G010.6311$-$00.3864	&	2	&	18:10:30.26	&	$-$19:55:30.0	&	15.1	&	27.8	&	35.5	&	15.4	&	2.0	&	$-$3.49	&	5.07	&	0.25	& &	1.6	&	$-$4.84	&	6.41	& &	2.0	&	$-$4.95	&	7.56	&	y	\\
G010.8411$-$02.5919	&	1	&	18:19:12.16	&	$-$20:47:32.2	&	6.1	&	23.7	&	28.5	&	15.1	&	1.0	&	12.27	&	1.91	&	0.41	& &	0.7	&	12.25	&	2.16	& &	0.4	&	12.33	&	2.36	&	y	\\
G010.8856+00.1221	&	1	&	18:09:08.13	&	$-$19:27:23.0	&	4.0	&	20.1	&	23.0	&	15.3	&	0.6	&	19.67	&	2.13	&	0.77	& &	0.3	&	19.69	&	2.51	& &	0.1	&	19.84	&	3.70	&	y	\\
G010.9592+00.0217	&	2	&	18:09:39.57	&	$-$19:26:25.8	&	6.5	&	23.8	&	28.8	&	15.1	&	0.9	&	21.41	&	2.90	&	0.29	& &	0.6	&	21.61	&	3.51	& &	0.4	&	21.20	&	4.24	&	y	\\
G010.9657+00.0083$^\dagger$	&	2	&	18:09:43.36	&	$-$19:26:28.6	&	$\cdots$	&	$\cdots$	&	$\cdots$	&	$\cdots$	&	0.2	&	19.16	&	3.98	&	$\cdots$	& &	0.2	&	18.36	&	3.20	& &	0.1	&	19.42	&	1.87	&	y	\\
G011.1109$-$00.4001	&	2	&	18:11:32.27	&	$-$19:30:40.6	&	5.1	&	19.1	&	21.6	&	15.7	&	1.6	&	0.12	&	3.04	&	1.33	& &	0.9	&	0.01	&	3.68	& &	0.5	&	$-$0.14	&	4.30	&	n	\\
G011.1723$-$00.0656	&	2	&	18:10:25.22	&	$-$19:17:46.3	&	$\cdots$	&	$\cdots$	&	$\cdots$	&	$\cdots$	&	$<$0.2	&	$\cdots$	&	$\cdots$	&	$\cdots$	& &	$\cdots$	&	$\cdots$	&	$\cdots$	& &	$\cdots$	&	$\cdots$	&	$\cdots$	&	n	\\
G011.3252$-$01.8040$^\dagger$	&	4	&	18:17:12.89	&	$-$19:59:36.9	&	$\cdots$	&	$\cdots$	&	$\cdots$	&	$\cdots$	&	0.4	&	9.00	&	0.95	&	$\cdots$	& &	$\cdots$	&	$\cdots$	&	$\cdots$	& &	$\cdots$	&	$\cdots$	&	$\cdots$	&	y	\\
G011.3757$-$01.6770$^\dagger$	&	2	&	18:16:50.47	&	$-$19:53:20.4	&	$\cdots$	&	$\cdots$	&	$\cdots$	&	$\cdots$	&	0.2	&	14.62	&	2.06	&	$\cdots$	& &	0.1	&	14.49	&	1.74	& &	$\cdots$	&	$\cdots$	&	$\cdots$	&	n	\\
G011.3757$-$01.6770$^\dagger$	&	2	&	18:16:50.47	&	$-$19:53:20.4	&	$\cdots$	&	$\cdots$	&	$\cdots$	&	$\cdots$	&	0.3	&	7.68	&	1.01	&	$\cdots$	& &	0.2	&	7.76	&	0.96	& &	$\cdots$	&	$\cdots$	&	$\cdots$	&	n	\\
G011.4201$-$01.6815	&	3	&	18:16:56.86	&	$-$19:51:07.2	&	6.0	&	17.8	&	19.9	&	15.5	&	2.5	&	9.18	&	1.88	&	1.50	& &	1.3	&	9.20	&	2.22	& &	0.4	&	9.23	&	2.56	&	n	\\
G011.5001$-$01.4857	&	3	&	18:16:22.58	&	$-$19:41:19.3	&	8.2	&	21.4	&	24.9	&	14.9	&	1.4	&	10.40	&	1.74	&	0.33	& &	0.8	&	10.41	&	1.97	& &	0.2	&	10.24	&	3.26	&	y	\\
G011.9019+00.7265$^\dagger$	&	1	&	18:08:58.89	&	$-$18:16:28.2	&	$\cdots$	&	$\cdots$	&	$\cdots$	&	$\cdots$	&	0.2	&	24.28	&	1.77	&	$\cdots$	& &	0.2	&	24.46	&	1.69	& &	$\cdots$	&	$\cdots$	&	$\cdots$	&	n	\\
G011.9454$-$00.0373	&	2	&	18:11:53.47	&	$-$18:36:18.3	&	$\cdots$	&	$\cdots$	&	$\cdots$	&	$\cdots$	&	$<$0.2	&	$\cdots$	&	$\cdots$	&	$\cdots$	& &	$\cdots$	&	$\cdots$	&	$\cdots$	& &	$\cdots$	&	$\cdots$	&	$\cdots$	&	n	\\
G012.0260$-$00.0317	&	1	&	18:12:02.04	&	$-$18:31:54.1	&	3.7	&	18.7	&	21.1	&	15.6	&	0.7	&	110.85	&	1.97	&	1.59	& &	0.4	&	110.81	&	2.50	& &	0.2	&	110.56	&	3.41	&	n	\\
G012.1993$-$00.0342	&	3	&	18:12:23.66	&	$-$18:22:51.6	&	4.6	&	19.5	&	22.2	&	15.7	&	1.3	&	50.86	&	2.62	&	1.52	& &	0.8	&	50.88	&	3.21	& &	0.5	&	51.65	&	4.71	&	y	\\
G012.4314$-$01.1117	&	2	&	18:16:51.33	&	$-$18:41:30.1	&	5.9	&	21.4	&	24.9	&	15.4	&	1.9	&	39.73	&	1.97	&	0.99	& &	1.2	&	39.68	&	2.44	& &	0.9	&	39.57	&	3.20	&	y	\\
G012.5932$-$00.5708$^\dagger$	&	4	&	18:15:10.43	&	$-$18:17:30.8	&	$\cdots$	&	$\cdots$	&	$\cdots$	&	$\cdots$	&	0.2	&	15.87	&	1.60	&	$\cdots$	& &	$\cdots$	&	$\cdots$	&	$\cdots$	& &	$\cdots$	&	$\cdots$	&	$\cdots$	&	n	\\
G012.8600$-$00.2737	&	2	&	18:14:36.67	&	$-$17:54:56.5	&	5.7	&	17.8	&	19.9	&	15.8	&	2.3	&	36.77	&	2.15	&	2.34	& &	1.4	&	36.69	&	2.65	& &	0.6	&	36.60	&	2.76	&	n	\\
G012.8909+00.4938	&	3	&	18:11:50.69	&	$-$17:31:14.5	&	6.1	&	20.1	&	23.0	&	16.0	&	2.3	&	32.92	&	3.22	&	2.23	& &	1.6	&	32.83	&	3.87	& &	1.0	&	32.81	&	4.24	&	y	\\
G012.9090$-$00.2607	&	1	&	18:14:39.71	&	$-$17:51:59.0	&	8.5	&	21.9	&	25.7	&	15.9	&	4.0	&	36.73	&	3.54	&	1.55	& &	2.8	&	36.88	&	3.68	& &	2.0	&	36.89	&	4.55	&	y	\\
G013.0105$-$00.1267	&	4	&	18:14:22.25	&	$-$17:42:47.8	&	3.8	&	$\cdots$	&	$\cdots$	&	$\cdots$	&	0.4	&	10.42	&	3.23	&	0.85	& &	$\cdots$	&	$\cdots$	&	$\cdots$	& &	$\cdots$	&	$\cdots$	&	$\cdots$	&	n	\\
G013.0105$-$00.1267	&	4	&	18:14:22.25	&	$-$17:42:47.8	&	3.8	&	$\cdots$	&	$\cdots$	&	$\cdots$	&	0.4	&	10.43	&	3.20	&	0.82	& &	$\cdots$	&	$\cdots$	&	$\cdots$	& &	$\cdots$	&	$\cdots$	&	$\cdots$	&	y	\\
G013.1840$-$00.1069	&	3	&	18:14:38.76	&	$-$17:33:05.0	&	5.2	&	16.6	&	18.2	&	15.7	&	2.1	&	53.39	&	1.61	&	2.83	& &	1.2	&	53.38	&	2.07	& &	0.6	&	53.18	&	3.60	&	n	\\
G013.1840$-$00.1069	&	3	&	18:14:38.76	&	$-$17:33:05.0	&	4.1	&	$\cdots$	&	$\cdots$	&	$\cdots$	&	1.0	&	36.79	&	0.86	&	1.45	& &	$\cdots$	&	$\cdots$	&	$\cdots$	& &	$\cdots$	&	$\cdots$	&	$\cdots$	&	n	\\
G013.3310$-$00.0407	&	1	&	18:14:41.78	&	$-$17:23:26.1	&	5.9	&	15.4	&	16.7	&	15.6	&	2.7	&	54.68	&	1.44	&	2.39	& &	1.3	&	54.65	&	1.87	& &	0.4	&	54.51	&	3.79	&	n	\\
G013.6562$-$00.5997	&	1	&	18:17:24.36	&	$-$17:22:14.5	&	6.0	&	18.2	&	20.4	&	15.9	&	2.8	&	47.45	&	2.08	&	3.52	& &	2.1	&	47.49	&	2.84	& &	2.0	&	47.58	&	3.31	&	y	\\
G013.8885$-$00.4760	&	3	&	18:17:24.67	&	$-$17:06:27.3	&	5.6	&	$\cdots$	&	$\cdots$	&	$\cdots$	&	0.7	&	22.68	&	0.92	&	0.31	& &	$\cdots$	&	$\cdots$	&	$\cdots$	& &	$\cdots$	&	$\cdots$	&	$\cdots$	&	n	\\
G014.0329$-$00.5155	&	1	&	18:17:50.61	&	$-$16:59:57.1	&	5.7	&	20.9	&	24.3	&	14.8	&	0.6	&	20.34	&	1.47	&	0.31	& &	0.3	&	20.42	&	1.39	& &	0.1	&	20.63	&	2.12	&	n	\\
G014.2166$-$00.6344	&	1	&	18:18:38.70	&	$-$16:53:37.3	&	3.8	&	14.6	&	15.7	&	15.2	&	0.6	&	19.75	&	1.36	&	1.10	& &	0.2	&	19.74	&	1.78	& &	$\cdots$	&	$\cdots$	&	$\cdots$	&	y	\\
G014.3313$-$00.6397	&	2	&	18:18:53.49	&	$-$16:47:42.7	&	7.2	&	20.8	&	24.1	&	15.8	&	2.8	&	22.26	&	2.87	&	1.55	& &	1.9	&	22.31	&	3.23	& &	1.5	&	22.65	&	3.55	&	y	\\
G014.4335$-$00.6969	&	1	&	18:19:18.23	&	$-$16:43:55.9	&	5.2	&	18.6	&	21.0	&	15.1	&	1.2	&	17.50	&	1.38	&	0.76	& &	0.6	&	17.65	&	1.66	& &	0.2	&	17.96	&	1.97	&	n	\\
G014.4886+00.0219	&	2	&	18:16:46.27	&	$-$16:20:34.4	&	3.5	&	17.3	&	19.1	&	15.8	&	0.5	&	22.88	&	3.01	&	1.70	& &	0.3	&	22.90	&	4.05	& &	0.1	&	22.69	&	4.13	&	n	\\
G014.5982+00.0202	&	2	&	18:16:59.68	&	$-$16:14:50.6	&	4.1	&	19.9	&	22.8	&	15.6	&	1.0	&	26.62	&	1.90	&	1.48	& &	0.6	&	26.56	&	2.53	& &	0.4	&	26.72	&	4.64	&	y	\\
G014.6087+00.0127	&	1	&	18:17:02.56	&	$-$16:14:30.1	&	5.1	&	20.1	&	23.1	&	15.9	&	1.8	&	24.63	&	2.92	&	2.02	& &	1.3	&	24.63	&	3.64	& &	1.1	&	24.60	&	4.50	&	y	\\
G014.9790$-$00.6649	&	3	&	18:20:15.60	&	$-$16:14:10.3	&	3.7	&	20.9	&	24.2	&	15.0	&	0.6	&	18.91	&	0.76	&	1.04	& &	0.3	&	18.83	&	1.17	& &	0.2	&	19.09	&	1.70	&	n	\\
G014.9958$-$00.6732	&	1	&	18:20:19.43	&	$-$16:13:31.0	&	4.8	&	23.1	&	27.7	&	16.0	&	1.1	&	19.43	&	4.79	&	1.34	& &	0.8	&	19.22	&	5.28	& &	0.5	&	19.07	&	5.04	&	y	\\
G015.0357$-$00.6795$^\dagger$	&	2	&	18:20:25.51	&	$-$16:11:35.5	&	$\cdots$	&	$\cdots$	&	$\cdots$	&	$\cdots$	&	0.4	&	19.20	&	2.59	&	$\cdots$	& &	0.5	&	19.40	&	2.85	& &	0.6	&	19.70	&	2.78	&	y	\\
G015.0939+00.1913$^\dagger$	&	1	&	18:17:20.86	&	$-$15:43:47.2	&	$\cdots$	&	$\cdots$	&	$\cdots$	&	$\cdots$	&	0.1	&	29.93	&	2.02	&	$\cdots$	& &	$\cdots$	&	$\cdots$	&	$\cdots$	& &	$\cdots$	&	$\cdots$	&	$\cdots$	&	y	\\
G015.1288$-$00.6717	&	1	&	18:20:34.75	&	$-$16:06:26.2	&	3.1	&	24.8	&	30.3	&	15.5	&	0.3	&	18.99	&	1.17	&	1.63	& &	0.2	&	18.80	&	1.74	& &	0.1	&	17.67	&	3.43	&	y	\\
G016.1438+00.0074	&	2	&	18:20:04.77	&	$-$14:53:31.2	&	3.9	&	16.2	&	17.7	&	15.5	&	0.9	&	44.84	&	1.84	&	1.54	& &	0.4	&	44.86	&	2.46	& &	0.2	&	44.90	&	3.75	&	y	\\
G016.7122+01.3119	&	1	&	18:16:26.85	&	$-$13:46:24.9	&	5.8	&	15.2	&	16.5	&	15.1	&	1.8	&	20.31	&	1.12	&	0.93	& &	0.6	&	20.26	&	1.39	& &	$\cdots$	&	$\cdots$	&	$\cdots$	&	n	\\
G016.7981+00.1264	&	1	&	18:20:55.27	&	$-$14:15:32.0	&	4.2	&	14.1	&	15.1	&	15.3	&	1.2	&	15.19	&	0.71	&	2.44	& &	0.5	&	15.19	&	0.88	& &	$\cdots$	&	$\cdots$	&	$\cdots$	&	n	\\
G016.8055+00.8149	&	1	&	18:18:25.96	&	$-$13:55:37.9	&	5.5	&	13.0	&	13.7	&	14.9	&	2.0	&	19.73	&	0.54	&	1.38	& &	0.5	&	19.75	&	0.70	& &	$\cdots$	&	$\cdots$	&	$\cdots$	&	y	\\
G016.8689$-$02.1552	&	1	&	18:29:24.26	&	$-$15:15:42.1	&	8.0	&	18.8	&	21.2	&	16.0	&	4.4	&	18.58	&	2.99	&	2.44	& &	3.0	&	18.54	&	3.82	& &	2.0	&	18.49	&	5.03	&	y	\\
\hline
\end{tabular}

$^a$ A superscript dagger symbol appended to the MSX name indicate source detections where we have been unable to fit the hyperfine satellite features of the NH$_3$ (1,1) emission.\\
$^b$ Source types are as follows: (1) H{\sc ii} regions; (2) YSOs; (3) H{\sc ii}/YSOs; (4) Young/Old stars.
$^c$ For non-detections we give a 3$\sigma$ upper limit.

Notes: Only a small portion of the data is provided here, the full table is only  available in electronic form at the CDS via anonymous ftp to cdsarc.u-strasbg.fr (130.79.125.5) or via http://cdsweb.u-strasbg.fr/cgi-bin/qcat?J/A+A/.
\end{minipage}
\end{table}

\end{landscape}

\begin{figure}
\begin{center}
\includegraphics[width=0.49\textwidth, trim= 0 0 0 0]{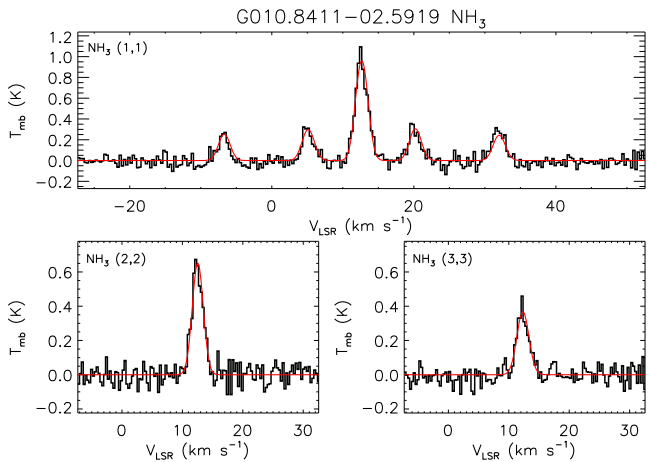}

\caption{\label{fig:nh3_spectra} Example of the NH$_3$ emission detected towards a young massive star, with NH$_3$ (1,1), (2,2) and (3,3) spectra on the top, lower left and lower right plots respectively. The model fits to these data are shown in red. The full version of this figure includes spectra taken towards all 591 sources and is only available in electronic form at the CDS via anonymous ftp to cdsarc.u-strasbg.fr (130.79.125.5) or via http://cdsweb.u-strasbg.fr/cgi-bin/qcat?J/A+A/.} 

\end{center}
\end{figure}

\subsection{NH$_3$: Molecular gas properties}
\label{sect:nh3_properties}

An example of the reduced spectra of the NH$_3$ (1,1), (2,2) and (3,3) inversion transitions and the fit to the data is presented in  Fig.\,\ref{fig:nh3_spectra}. We detect NH$_3$ (1,1) emission towards $\sim$80 of the sample with the hyperfine structure seen towards the majority ($\sim$65\,per\,cent) of the observed sources, and used an IDL routine to simultaneously fit all 18 hyperfine components and to derive the optical depth and line widths. The hyperfine structure is generally too weak to be observed in the NH$_3$ (2,2) and (3,3) transitions: for these lines (as well as the remaining NH$_3$ (1,1) lines) we have obtained corrected antenna temperatures by fitting a single Gaussian profile to the main line. In all cases the NH$_3$ (1,1) and (2,2) line widths have been obtained by fitting the hyperfine components to their respective main line emission to remove the effects of line broadening due to optical depth. The resulting fits to the data are shown in Fig.\,\ref{fig:nh3_spectra} and the fitted parameters, velocity, line width, optical depth and corrected antenna temperature  are presented in Table\,\ref{tbl:nh3_parameters}.\footnote{The full version of  Fig.\,\ref{fig:nh3_spectra} and Table\,\ref{tbl:nh3_parameters} are only available in electronic form at the CDS via anonymous ftp to cdsarc.u-strasbg.fr (130.79.125.5) or via http://cdsweb.u-strasbg.fr/cgi-bin/qcat?J/A+A/.}

\begin{figure}
\begin{center}
\includegraphics[width=0.23\textwidth, trim= 30 0 10 20]{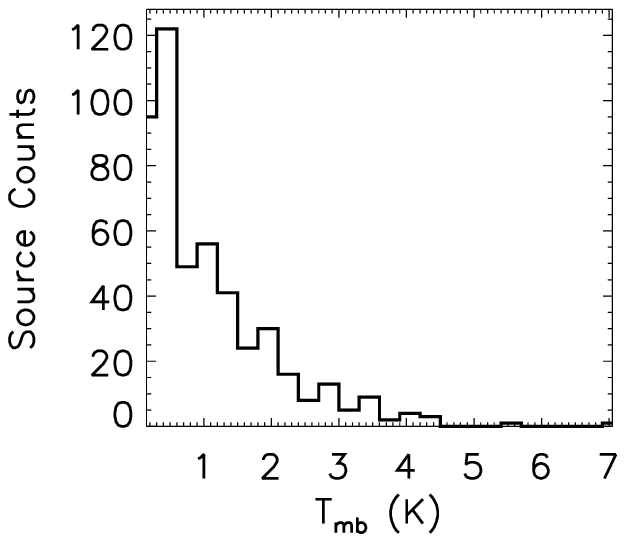}
\includegraphics[width=0.23\textwidth, trim= 30 0 10 20]{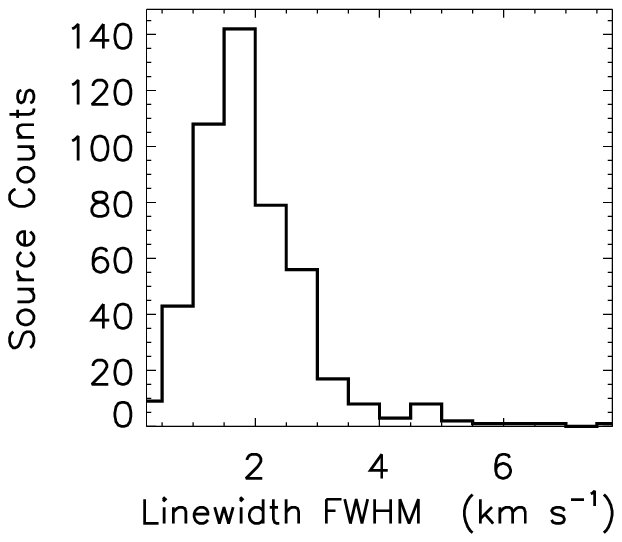}
\includegraphics[width=0.23\textwidth, trim= 30 0 10 20]{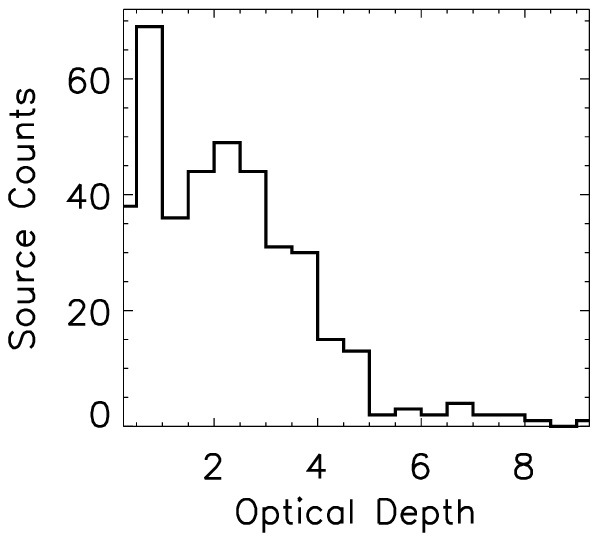}
\includegraphics[width=0.23\textwidth, trim= 30 0 10 20]{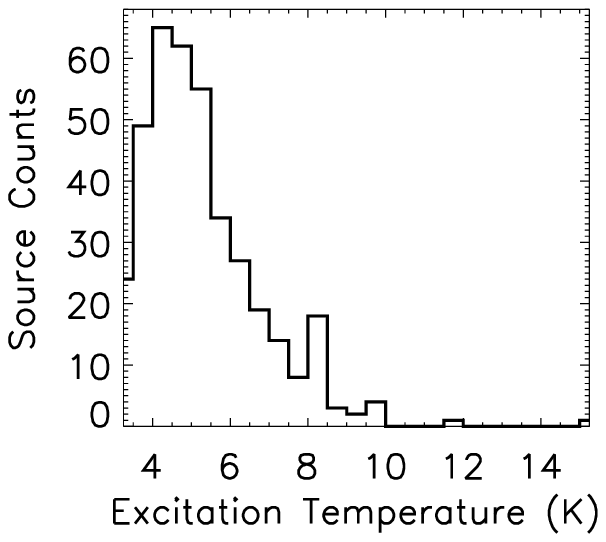}
\includegraphics[width=0.23\textwidth, trim= 30 0 10 20]{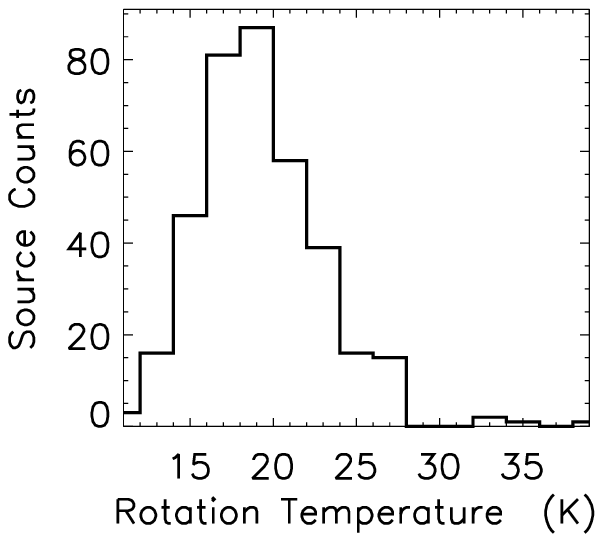}
\includegraphics[width=0.23\textwidth, trim= 30 0 10 20]{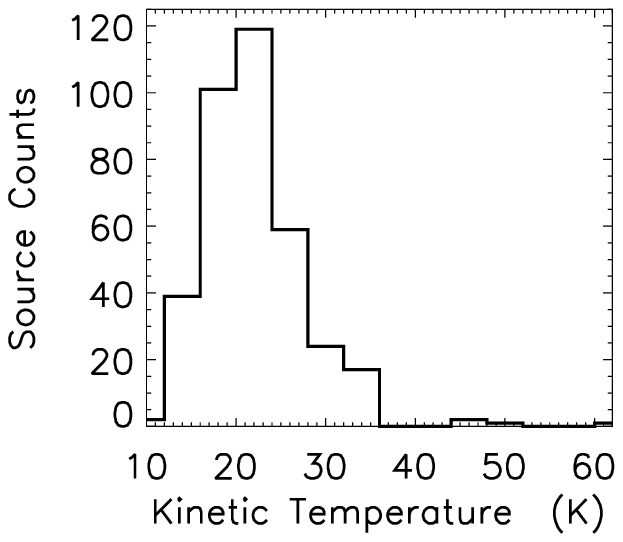}
\includegraphics[width=0.23\textwidth, trim= 30 0 10 20]{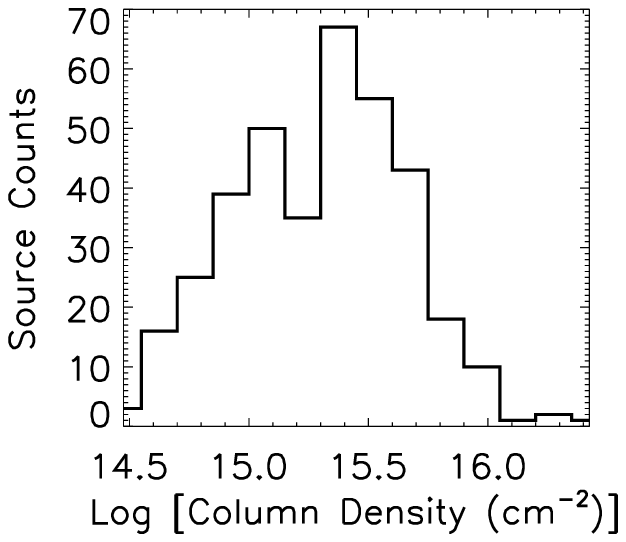}
\includegraphics[width=0.23\textwidth, trim= 30 0 10 20]{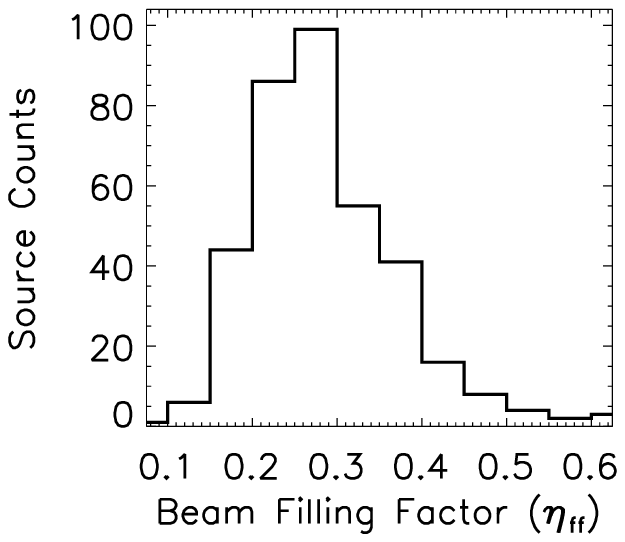}

\caption{\label{fig:nh3_dist} Distributions of measured and derived parameters for the observed sample. The mean and standard derivation, and maximum and minimum values of each distribution are presented in Table\,\ref{tbl:derived_para}.} 

\end{center}
\end{figure}

\begin{table*}

\begin{center}\caption{Summary of derived parameters}
\label{tbl:derived_para}
\begin{minipage}{\linewidth}
\small
\begin{tabular}{lc.....}
\hline \hline
  \multicolumn{1}{c}{Parameter}&  \multicolumn{1}{c}{Number}&	\multicolumn{1}{c}{Mean}  &	\multicolumn{1}{c}{Median} &\multicolumn{1}{c}{$1\sigma$} & \multicolumn{1}{c}{Min}& \multicolumn{1}{c}{Max}\\
\hline
$T_{\rm{mb}}$ (K)								&	479 	& 1.1	& 0.7	& 1.0		&	0.1			&	6.9		\\
$\Delta$V (\kms)								&	479 	& 1.9	& 1.7	& 0.9		&	0.2			&	7.8		\\
$\tau_{(1,1)}$ 									&	386 	& 2.2	& 2.1	& 1.6		&	0.3			&	9.2		\\
$T_{\rm{ex}}$ (K)								&	386 	& 5.3	& 4.9	& 1.5		&	3.1			&	15.1		\\
$T_{\rm{rot}}$ (K)								&	365 	& 19.3	& 19.2	& 3.8		&	10.2		&	39.0		\\
$T_{\rm{kin}}$ (K)								&	365 	& 22.1	& 21.4	& 5.9		&	10.5		&	60.8		\\
Log[N(NH$_3$) (cm$^{-2}$)]						&	365 	& 15.3	& 15.3	& 0.4		&	14.4		&	16.5		\\
Beam$_{\rm{ff}}$ 								&	365 	& 0.28	& 0.27	& 0.09		&	0.09		&	0.63	\\
\hline
Log[S$_{\nu{\rm{(H_2O)}}}$ (Jy)]				&	298 	& 1.1	& 0.9 & 1.0		&	-0.5		&	5.8		\\
Log[$\int S_{\nu{\rm{(H_2O)}}}$ (Jy\,\kms)]		&	298 	& 1.3	& 1.2	&1.1		&	-1.7		&	6.7		\\
H$_2$O Vel. Range (\kms)						&	298 	& 24.7	& 15.2	& 32.5		&	0.00		&	362.0		\\
Log[L$_{\rm{(H_2O)}}$ (\lsun)]					&	298 	& -5.0	& -5.0	& 1.3		&	-8.9		&	1.2		\\
\hline\\
\end{tabular}\\

\end{minipage}

\end{center}
\end{table*}

We obtain the total optical depth ($\tau_{\rm{(1,1)}}$) of the NH$_3$ (1,1) transition from the hyperfine fitting performed. If we assume that the optical depth ratios between the hyperfine transitions are equal to the ratios of the theoretical transition intensities \citep{rydbeck1977}, then the optical depth of the NH$_3$ (1,1) main quadrupole transition ($\tau_{\rm{(1,1,m)}}$) is approximately half that of the total optical depth. We can now estimate the excitation temperature ($T_{\rm{ex}}$) using the detection equation assuming the gas is in local thermal equilibrium (LTE), a background temperature of 2.73\,K and a beam filling factor ($\eta_{\rm{ff}}$) of unity; the beam filling factor is a measure of the fraction of the telescope beam that is filled by the source. Using the main line optical depth ($\tau_{\rm{(1,1,m)}}$) of the NH$_3$ (1,1) transition, we obtain the excitation temperature using:

\begin{equation}
T_{\mathrm{A}}^{*} = \eta_{\mathrm{mb}} \eta_{\mathrm{ff}} [J(T_{\mathrm{ex}})-J(T_{\mathrm{bg}})][1-e^{-\tau_{\rm{(1,1,m)}}}],
\end{equation}

\noindent where,

\begin{equation}
J_\nu(T) =\frac{\mathrm{h} \nu/\mathrm{k}}{(e^{\mathrm{h} \nu/\mathrm{k} T}-1)}. 
\end{equation}

The rotation temperature ($T_{\rm{rot}}$) associated with the NH$_3$ (2,2) and (1,1) transitions can be calculated using the line intensities and the (1,1) main quadrupole transition optical depth  \citep{ho1983}:

\begin{equation}
T_\mathrm{rot} = \frac{-T_0}{\ln \left\{\frac{-0.282}{\tau_(1,1,{\rm{m}})} \ln \left[1-\frac{ T_{\mathrm{A}(2,2,{\rm{m}})}^{*}}{T_{\mathrm{A}(1,1,{\rm{m}})}^{*}}\left(1-\mathrm{e}^{-\tau_{(1,1,{\rm{m}})}} \right) \right] \right\}}~[{\rm{K}}],
\end{equation}

\noindent where $T_0 = \frac{E_{(2,2)} - E_{(1,1)}}{\mathrm{k_B}} \approx 41.5$\,K is the temperature associated with the energy difference between (1,1) and (2,2) levels, and $ T_{\mathrm{A}(1,1,\mathrm{m})}^{*}$ and $T_{\mathrm{A}(2,2,\mathrm{m})}^{*}$ are the measured corrected peak antenna temperatures of the main quadrupole transition of the NH$_3$ (1,1) and (2,2) lines, respectively. The next step is to relate the rotation temperature to the kinetic temperature of the gas. Empirical results reveal that at low temperatures ($<15$\,K) the rotation and kinetic temperature are approximately equivalent; however, they begin to deviate at higher temperature and thus analytic expressions underestimate the rotation temperatures for kinetic temperatures above 40\,K (\citealt{walmsley1983,ho1983}). For \Tk\ $< T_0$, a relationship between excitation and kinetic temperature may be calculated by consideration of (1,1), (2,2), and (2,1) states only \citep{swift2005, walmsley1983}, such that:

\begin{equation}
T_\mathrm{rot} = \frac{T_\mathrm{kin}} {1+\frac{T_\mathrm{kin}}{T_0}\ln\left[ 1+0.6\times \exp\left(\frac{-15.7}{T_\mathrm{kin}}\right)\right]}~[{\rm{K}}]
\end{equation}

The kinetic temperatures calculated using Eqn.\,3 may be overestimated for sources where $T_\mathrm{rot}$ $\simeq T_0$. The mean rotation temperature is $\sim$20\,K and so this is not a concern for the vast majority of our detections. However, we do find a handful of sources where this might be an issue (e.g., G019.6085$-$00.2357 and G049.4903$-$00.3694).

If excitation conditions are homogeneous along the beam and all hyperfine lines have the same excitation temperature, then the column densities at a given $(J,K=J)$ transition can be written as a function of the \emph{total} column density \citep{mangum1992}:

\begin{equation}\label{eq:coldensity}
N_{\rm{(1,1)}} = 6.60 \times 10^{14} \Delta{\rm{V}}_{\rm{(1,1)}}\tau_{\rm{(1,1,m)}}\frac{T_\mathrm{rot}}{\nu_{\rm{(1,1)}}} ~~ [{\rm{cm}}^{-2}]
\end{equation}

\noindent where $N_{\rm{(1,1)}}$ is the column density of the NH$_3$ (1,1) transition, $\Delta{\rm{V}}_{\rm{(1,1)}}$ is the FWHM line width of the NH$_3$ (1,1) transition in \kms\ and $\nu_{\rm{(1,1)}}$ is the transition frequency in GHz. Finally we estimate the total ammonia column density following \citet{li2003}:

\begin{equation}\label{eq:coldensity_total}
N_{\rm{NH_3}} = N_{\rm{(1,1)}} \left[ 1+\frac{1}{3}{\rm{exp}}\left(\frac{23.1}{T_\mathrm{rot}}\right)+\frac{5}{3}{\rm{exp}}\left(\frac{-41.2}{T_\mathrm{rot}}\right)+\frac{14}{3}{\rm{exp}}\left(\frac{-99.4}{T_\mathrm{rot}}\right)  \right] ~~ [{\rm{cm}}^{-2}]
\end{equation}

In Fig.\,\ref{fig:nh3_dist} we present histograms showing the distributions of the various parameters derived in this subsection and present a summary of the derived parameters in Table\,\ref{tbl:derived_para}.

\subsection{Beam filling factor}
\label{sect:filling_factor}

\begin{figure}
\begin{center}
\includegraphics[width=0.49\textwidth, trim= 0 0 0 0]{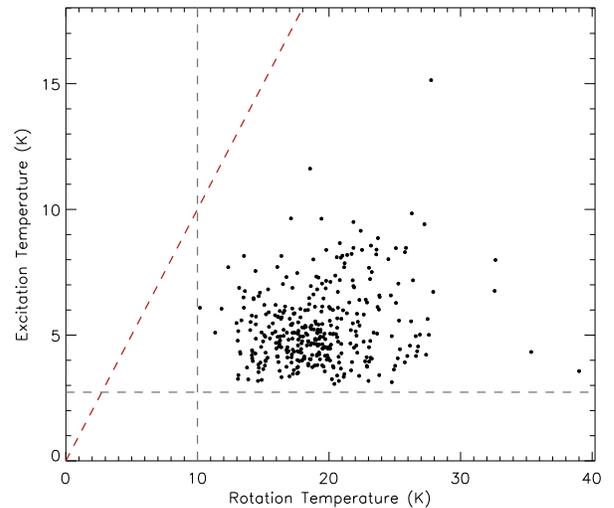}

\caption{\label{fig:tex_vs_trot} Plot comparing the derived excitation and rotation temperatures. The red dashed line indicates the line of equality. The black vertical line indicates the minimum cloud temperatures expected due to heating by cosmic rays and the far-uv radiation field ($\sim$10\,K; \citealt{evans1999}), whilst the black horizontal line indicates the temperature of the cosmic microwave background radiation.} 

\end{center}
\end{figure}

In Eqn.\,1 presented in the previous subsection we assumed the beam filling factor to be of order unity; however, having calculated the excitation and rotation temperatures of the molecular gas we are in a position to test this assumption. In Fig.\,\ref{fig:tex_vs_trot} we present a scatter plot comparing the derived excitation and rotation temperatures; the dashed line indicates the line of equality where $T_{\rm{ex}}=T_{\rm{rot}}$. This plot clearly shows that the rotation temperatures are systematically $\sim$4 times higher than the excitation temperatures. The excitation temperatures are calculated from a single transition and assume the beam is uniformly filled (i.e., $\eta_{\rm{ff}} \sim 1$). The assumption that the beam is uniformly filled results in unfeasibly low inversion transition excitation temperatures (typically $\sim$5\,K). The rotation temperature is derived from the line intensity ratio of the NH$_3$ (1,1) and (2,2) transitions and therefore the beam filling factor is effectively divided out and thus provides a more reliable estimate of the gas temperature.

The assumption of LTE includes the inversion transitions and implies that $T_{\rm{ex}}$ should be equal to $T_{\rm{rot}}$. Hence, our derived values of  $T_{\rm{ex}}$ are actually $\simeq$ $T_{\rm{rot}} \times \eta_{\rm{ff}}$ and we can estimate the beam filling factor from the ratio of the excitation and rotation temperatures, i.e., $\eta_{\rm{ff}}=T_{\rm{ex}}/T_{\rm{rot}}$. We note the non-linear nature of Eqn.\,1, however, given that $J(T) \sim T$ in Eqn.\,2 for $T\gg h\nu/k$, which corresponds to a temperature of 1.1\,K at $\nu$= 23\,GHz and so is nearly always the case for these transitions, we find that Eqn.\,1 is approximately linear in $T$.  We derive minimum and maximum beam filling factors of 0.1 and 0.6 respectively, with a mean value of $\sim$0.3; this compare  well to the values reported by \citet{pillai2006} ($\eta_{\rm{ff}} \sim$0.3--0.5) from ammonia mapping observations of a small sample of infrared dark clouds (IRDCs) and \citet{Rosolowsky2008} ($\eta_{\rm{ff}} \sim$0.3) from GBT observations of dense cores in Perseus. The observations of \citet{pillai2006} were conducted with the Effelsberg 100-m telescope and have a similar resolution and sensitivity to the observations presented here. Since it is unlikely that the inversion transitions would be sub-thermally excited, whilst the rotation transitions are thermalised, the implication is that the low beam filling factors are real and so there is likely to be high-contrast substructure within the GBT beam.    

Further evidence of the presence of substructure comes from high-resolution observations of a sample of high-mass star forming clumps observed with the Compact Array reported by \citet{longmore2007}. These authors found typical clump sizes of $\sim$10-15\arcsec\ with multiple cores being detected within the 2\arcmin\ primary beam in approximately half of the fields observed. Using the beam filling factor we estimate the typical angular diameter to be similar to those reported by \citet{longmore2007}, which at the mean distance of our sample ($\sim$5\,kpc), corresponds to physical diameters of 0.2-0.4\,pc. These observations are therefore probing molecular clumps that are likely going to form single stars or small multiple stellar systems rather than whole clusters.

\begin{figure}
\begin{center}
\includegraphics[width=0.49\textwidth, trim= 0 0 0 0]{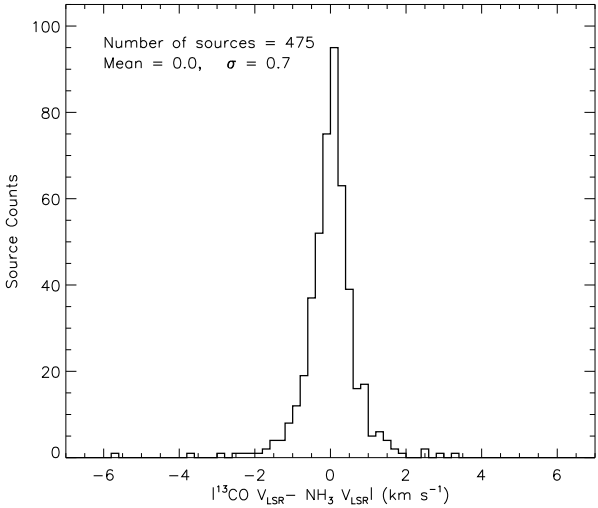}

\caption{\label{fig:co_nh3_vlsr} Distribution of the differences between the velocity assigned from $^{13}$CO data and the velocity of the NH$_3$ (1,1) inversion transition. In the upper left corner we give the statistical parameters of the distribution; these have been calculated for all sources where the absolute velocity difference is less than 5\,\kms. Note the $x$-axis has been truncated $\pm$7\,\kms; the details of the two sources that have velocity differences greater than this can be found in Table\,\ref{tbl:vlsr_diff}.} 

\end{center}
\end{figure}

\subsection{NH$_3$-CO velocity comparison}

The RMS project has previously used $^{13}$CO rotational transitions to determine source velocities (\citealt{urquhart_13co_south,urquhart_13co_north}) and kinematic distances, using the Galactic rotation curve of \citet{brand1993}. However, we detected multiple velocity components in $^{13}$CO towards approximately 60\,per\,cent of our sources. In the majority of these cases it was straightforward to identify the correct velocity component. However, there were a small number in which the velocity assignment was less reliable. With the NH$_3$ observations in hand we are in a position to check the previous velocity assignments and make corrections where necessary. In Fig.\,\ref{fig:co_nh3_vlsr} we present a histogram showing the difference between the assigned $^{13}$CO velocity and the NH$_3$ (1,1) velocity. A cursory inspection of this plot reveals the velocity assigned from the CO spectra is in excellent agreement with the  NH$_3$ velocities. Indeed there are only eleven sources where the velocities differ by more than 3$\sigma$ ($\sim$2.1\,\kms); this corresponds to approximately 2\,per\,cent of the sample. In Table\,\ref{tbl:vlsr_diff} we give the names and velocities for these eleven sources.

Inspection of the individual $^{13}$CO profiles (\citealt{urquhart_13co_north}) for all sources in Table\,\ref{tbl:vlsr_diff} with velocity differences less than 5\,\kms\ reveals the presence of two or more blended components, or a single emission feature with uncharacteristically broad line width of $\sim$5-7\,\kms, which probably indicates that a number of velocity components have been blended into a single feature. The differences in the CO and NH$_3$ velocities for these sources can easily be understood in terms of blended components along the line of sight. The remaining three sources have velocity differences greater than 5\,\kms; inspection of their CO spectra reveals the presence of several distinct emission features and, given that NH$_3$ is a tracer of dense gas, it is likely that the wrong CO component was assigned. We have used the NH$_3$ velocity to determine the kinematic properties of these three sources and have updated the RMS database accordingly.

\begin{table}
\begin{center}\caption{RMS sources where the differences between the assigned $^{13}$CO velocity and NH$_3$ (1,1) velocities are greater than 3$\sigma$ (i.e., $|\Delta{\rm{V}}| >2.1$\,\kms). }
\label{tbl:vlsr_diff}
\begin{minipage}{\linewidth}
\begin{tabular}{l...}
\hline \hline
  \multicolumn{1}{c}{MSX Name}&  \multicolumn{1}{c}{\vlsr($^{13}$CO)}&	\multicolumn{1}{c}{\vlsr(NH$_{3}$)}&\multicolumn{1}{c}{$|\Delta{\rm{V}}|$}\\
  & \multicolumn{1}{c}{(\kms)}&	\multicolumn{1}{c}{(\kms)}&	\multicolumn{1}{c}{(\kms)}\\
\hline
G031.4134+00.3092&	98.7&	96.3	&2.4\\
G010.4413+00.0101&	68.8&	66.4	&2.5\\
G035.1992$-$01.7424&	42.4	&45.0	&2.6\\
G016.9512+00.7806&	22.3	&24.9		&2.6\\
G045.4658+00.0457&	59.4&	62.1	&2.8\\
G014.4335$-$00.6969&	20.4	&17.5	&2.9\\
G030.7206$-$00.0826&	90.4	&93.7		&3.3\\
G015.1288$-$00.6717&	22.7	&19.0		&3.7\\
G038.2577$-$00.0733&	17.5	&11.9	&5.7\\
G030.1043$-$00.0738&	101.1	&88.3	&12.8\\
G017.1141$-$00.1130&	44.4	&93.1	&48.7\\
\hline\\
\end{tabular}\\

\end{minipage}

\end{center}
\end{table}

\subsection{Water maser properties}
\label{sect:Water_maser_properties}

\begin{figure}
\begin{center}
\includegraphics[width=0.49\textwidth, trim= 30 0 0 0]{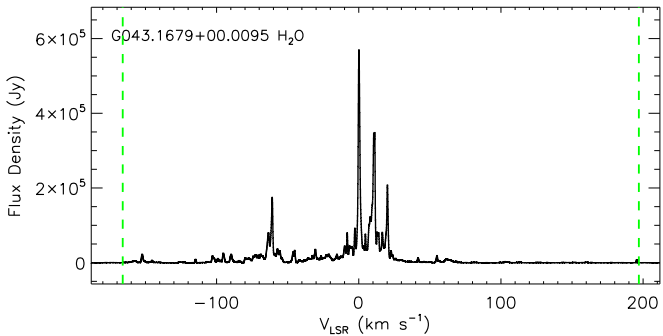}

\caption{\label{fig:h2o_spectra} Example of H$_2$O maser emission detected towards the MSX source G043.1679+00.0095. The vertical dashed lines indicate the minimum and maximum velocity ranges over which maser emission is detected. The full version of this figure includes spectra taken towards 597 sources and is only available in electronic form at the CDS via anonymous ftp to cdsarc.u-strasbg.fr (130.79.125.5) or via http://cdsweb.u-strasbg.fr/cgi-bin/qcat?J/A+A/.} 

\end{center}
\end{figure}

\begin{table*}

\begin{center}\caption{Water maser properties.}
\label{tbl:h2o_masers}
\begin{minipage}{\linewidth}
\begin{tabular}{l...........}
\hline \hline
  \multicolumn{1}{c}{MSX Name}&  \multicolumn{1}{c}{Type$^a$}&	\multicolumn{1}{c}{r.m.s.}&\multicolumn{1}{c}{$V_{\rm{min}}$}&	\multicolumn{1}{c}{$V_{\rm{max}}$}&	\multicolumn{1}{c}{$V_{\rm{peak}}$}&	\multicolumn{1}{c}{$V_{\rm{NH_3}}$}   &\multicolumn{1}{c}{Log[$S_\nu$]} & \multicolumn{1}{c}{Log[$\int S_\nu {\rm{d}}V$]}& \multicolumn{1}{c}{Log[L$_{\rm{(H_2O)}}$]}& \multicolumn{1}{c}{Log[L$_{\rm{bol}}$]$^b$}\\
  &  &	\multicolumn{1}{c}{(Jy beam$^{-1}$)}&\multicolumn{1}{c}{(\kms)}&	\multicolumn{1}{c}{(\kms)}&	\multicolumn{1}{c}{(\kms)}&	\multicolumn{1}{c}{(\kms)}   &\multicolumn{1}{c}{(Jy)} & \multicolumn{1}{c}{(Jy \kms)}& \multicolumn{1}{c}{(\lsun)}& \multicolumn{1}{c}{(\lsun)}\\
\hline
G229.5711+00.1525&2&0.10&40.4&60.9&46.7&52.8&1.80&2.08&-4.07&4.11\\G236.8158+01.9821&2&0.11&37.3&65.2&52.0&52.8&1.86&2.43&-3.76&3.80\\G038.3543$-$00.9519&2&0.09&11.2&13.0&12.3&16.9&0.57&0.44&-7.09&1.96\\G232.0766$-$02.2767&2&0.09&41.4&48.5&42.0&42.1&0.69&0.74&-5.64&3.71\\G220.4587$-$00.6081&2&0.09&23.6&33.6&30.5&29.6&1.02&1.04&-5.64&3.18\\G224.6075$-$01.0063&2&0.08&27.9&29.7&28.7&16.7&0.41&0.38&-6.86&3.19\\G079.8855+02.5517&2&0.10&-2.0&20.2&4.2&5.9&0.99&1.43&-5.75&3.55\\G079.8749+01.1821&1&0.10&-7.4&3.1&-4.2&-4.3&1.27&1.25&-5.93&3.83\\G082.0333+02.3249&1&0.09&-4.8&8.0&-0.8&0.3&1.20&1.55&-5.63&3.51\\G083.7071+03.2817&2&0.09&-7.1&8.4&-6.3&-3.6&0.59&0.86&-6.32&3.75\\G080.8624+00.3827&2&0.09&-9.6&4.2&-7.7&-1.9&1.36&1.64&-5.54&3.75\\G081.8652+00.7800&2&0.12&-28.5&46.5&10.6&9.4&3.22&3.81&-3.37&3.70\\G081.8789+00.7822&1&0.11&-13.2&20.9&10.6&8.1&2.73&3.30&-3.88&4.49\\G081.7131+00.5792&2&0.10&-7.8&15.2&1.4&-3.6&1.34&1.72&-5.45&3.83\\G081.7220+00.5699&1&0.10&-44.8&33.8&-5.5&-2.8&3.06&3.21&-3.97&2.80\\G081.7522+00.5906&2&0.10&-5.6&27.4&0.6&-4.0&0.84&1.21&-5.97&3.39\\G081.7133+00.5589&1&0.10&-6.6&7.7&-0.1&-3.8&0.31&0.85&-6.33&3.42\\G081.7624+00.5916&2&0.11&-14.8&32.9&-9.1&-4.4&0.51&0.96&-6.22&3.00\\G081.6632+00.4651&2&0.10&8.1&22.5&8.4&19.3&0.63&0.52&-6.66&2.54\\G084.1940+01.4388&2&0.10&-0.2&41.0&38.7&-1.8&0.56&0.89&-6.29&3.67\\G085.4102+00.0032&3&0.10&-39.6&-12.9&-32.5&-35.8&1.84&2.24&-3.74&4.50\\G084.9505$-$00.6910&2&0.09&-34.5&-29.6&-33.9&-34.9&0.20&0.18&-5.81&4.28\\G095.0531+03.9724&2&0.11&-91.9&-82.5&-88.4&-84.1&1.63&1.78&-3.79&4.28\\G093.1610+01.8687&2&0.11&-70.0&-70.0&-70.3&-63.1&0.72&0.36&-5.43&4.17\\G097.5268+03.1837&3&0.12&-111.5&-52.6&-76.1&-70.0&2.42&3.11&-2.65&4.69\\G094.2615$-$00.4116&2&0.10&-76.8&-35.6&-36.4&-46.6&0.77&1.04&-5.02&4.10\\G094.4637$-$00.8043&2&0.10&-48.6&-43.5&-46.2&-44.6&1.03&1.16&-4.90&4.50\\G096.4353+01.3233&3&0.11&-70.0&-62.5&-69.9&-68.9&-0.03&0.07&-5.68&4.27\\G094.6028$-$01.7966&2&0.12&-60.2&-39.8&-56.0&-43.9&2.43&2.88&-3.64&4.37\\G095.0026$-$01.5779&2&0.10&-46.6&-38.4&-45.8&-40.6&1.41&1.39&-5.13&3.48\\
\hline\\
\end{tabular}\\
$^a$ Source types are as follows: (1) H{\sc ii} regions; (2) YSOs; (3) H{\sc ii}/YSOs; (4) Young/Old stars.

$^b$ Bolometric luminosities have been calculated using the integrated fluxes derived by \citet{mottram2011a}.\\
Notes: Only a small portion of the data is provided here, the full table is only  available in electronic form at the CDS via anonymous ftp to cdsarc.u-strasbg.fr (130.79.125.5) or via http://cdsweb.u-strasbg.fr/cgi-bin/qcat?J/A+A/.
\end{minipage}

\end{center}
\end{table*}

\begin{figure}
\begin{center}
\includegraphics[width=0.23\textwidth, trim= 30 0 0 0]{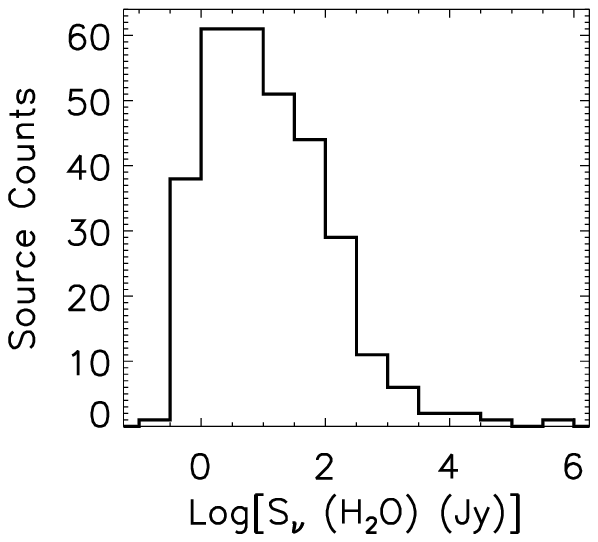}
\includegraphics[width=0.23\textwidth, trim= 30 0 0 0]{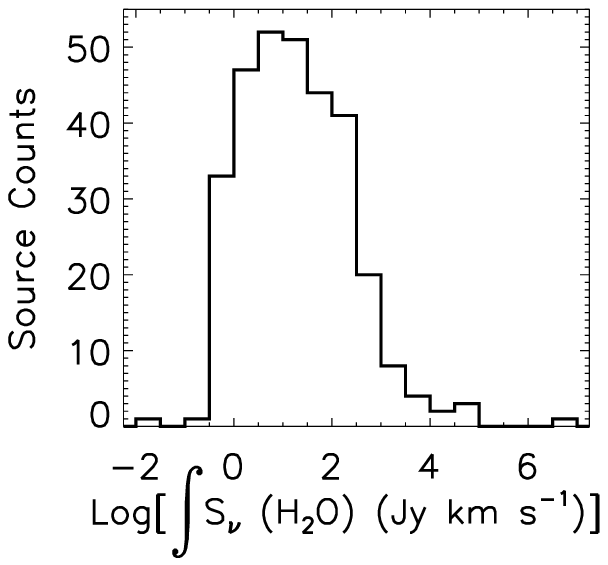}
\includegraphics[width=0.23\textwidth, trim= 30 0 0 0]{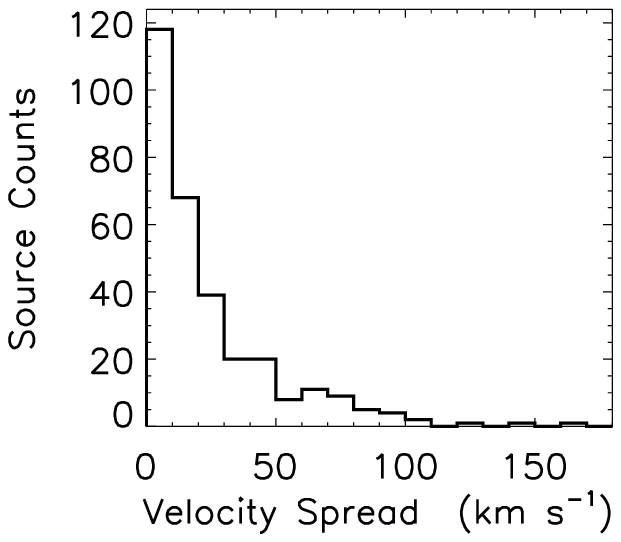}
\includegraphics[width=0.23\textwidth, trim= 30 0 0 0]{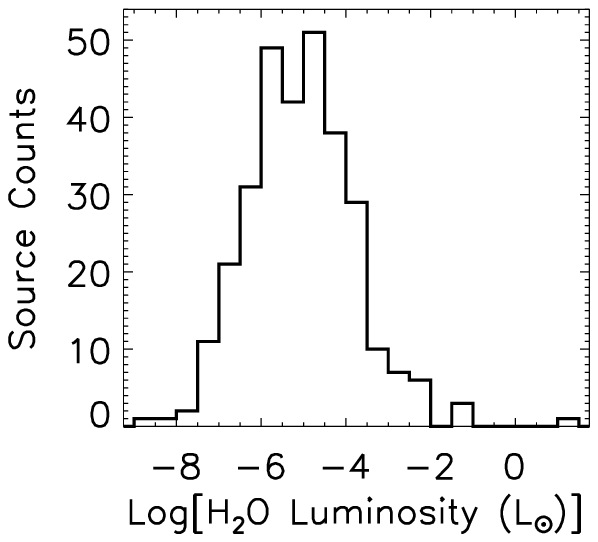}

\caption{\label{fig:h2o_dist}The distributions of derived water maser properties. The upper panels show the peak and integrated flux distribution (the bin size is  0.5\,dex). The lower panels show the total velocity range over which the maser spots are seen in each spectrum and the isotropic luminosities. For the velocity distribution a bin size of 10\,\kms\ has been used and the $x$-axis has been truncated at 150\,\kms\ --- only one source has a total velocity range larger than this. The luminosity distribution uses a bin size of 0.5\,dex.} 

\end{center}
\end{figure}

In Fig.\,\ref{fig:h2o_spectra} we present the water-maser spectrum observed towards the MSX source G043.1679+00.0095.\footnote{The full version of this figure, showing all 308 spectra, is available in electronic form at the CDS via anonymous ftp to cdsarc.u-strasbg.fr (130.79.125.5) or via http://cdsweb.u-strasbg.fr/cgi-bin/qcat?J/A+A/.} This particular source is an H{\sc ii} region associated with the W49A star-forming complex, and the water maser detected towards this region is the brightest we have found with a peak flux value of $\sim6 \times 10^5$\,Jy. Maser spectra often consist of a number of distinct emission peaks, usually referred to as maser spots, spread over a range of velocities. The vertical dashed lines in Fig.\,\ref{fig:h2o_spectra} show the minimum and maximum velocities over which maser spots are found above a 3\,$\sigma$ detection limit towards this object.

For all of the maser emission detected we measure the peak flux and velocity of the brightest maser spot, the integrated flux and the minimum and maximum velocities over which the maser spots are distributed.  These values are tabulated in Table\,\ref{tbl:h2o_masers}. In the upper panels of Fig.\,\ref{fig:h2o_dist} we present the peak and integrated flux-density distributions and in the lower left panel of this figure we present a histogram of the total velocity range over which the maser spots are seen in each spectrum. The median peak flux is $\sim$8\,Jy with minimum and maximum peak fluxes of 0.26 and 5.7$\times10^5$\,Jy respectively. The largest velocity range of maser spots is seen towards G043.1679+00.0095, which has a range of 362\,\kms. However, only six sources have total velocity ranges larger than 100\,\kms, and the majority ($\sim$60\,per cent) have total velocity ranges of 20\,\kms\ or less. In the lower right panel of Fig.\,\ref{fig:h2o_dist} we present the isotropic H$_2$O maser luminosity distribution which is discussed in the following paragraphs. A summary of minimum, maximum, mean, and median values  can be found in the last four rows of Table\,\ref{tbl:derived_para} for the four distributions presented in Fig.\,\ref{fig:h2o_dist}.

Kinematic distances have been determined from source velocities derived from $^{13}$CO (\citealt{urquhart_13co_north, urquhart_13co_south}), or the NH$_3$ observations presented here, using the Galactic rotation curve of \citet{brand1993}, assuming the distance to the Galactic centre is 8.5\,kpc and the angular velocity of the Sun is 220\,\kms. We have resolved the twofold kinematic distance ambiguity that affects sources located within the Solar circle using archival HI data (see \citealt{urquhart2011} for details). These kinematic distances have been adjusted for sources found to be associated with complexes that have well-determined distances and for sources for which maser parallax and photometric distances are available.

In total we have estimated distances to 573 of the RMS sources observed as part of this programme; this includes 295 sources towards which water maser emission is detected. We use these distances to estimate the isotropic luminosity for each water maser by integrating all of the emission in each channel above 3$\sigma$ and using the following equation taken from \citet{anglada1996}:

\begin{equation}\label{eq:h2o_luminosity}
\left[ \frac{L{({\rm{H_2O}})}}{{\rm{L}}_\odot} \right] = 2.30 \times 10^{-8} \left[ \frac{\int S\nu {\rm{d}}V} {{\rm{Jy\,km\,s^{-1}}}}\right]\left[\frac{D}{{\rm{kpc}}}\right]
\end{equation}

\noindent where $D$ is the distance to the source and the integral extends over all components of the spectrum over  3$\sigma$.  The estimated water maser and bolometric luminosities obtained using the bolometric fluxes calculated by \citet{mottram2011a} can be found in the last two columns of Table\,\ref{tbl:h2o_masers}. The water maser properties derived here will be discussed in detail in Sect.\,\ref{sect:h2o_correlations}.

\section{Results and analysis}
\label{sect:results}

\setlength{\tabcolsep}{4pt}

\begin{table*}

\begin{center}\caption{Observation and detection statistics.}
\label{tbl:detection_stats}
\begin{minipage}{\linewidth}
\begin{tabular}{lccccccccccccc}
\hline \hline
 & &  	\multicolumn{8}{c}{NH$_3$ Detection Statistics$^{\rm{a}}$}  &&\multicolumn{3}{c}{H$_2$O Detection Statistics} \\
\cline{2-4} \cline{5-10} \cline{12-14} 
 Source Type& Observed & Detected	 & Ratio &  \multicolumn{1}{c}{Hyperfine} & Ratio &(2,2) & Ratio&  (3,3) & Ratio  &&Observed &Detections & Ratio\\

\hline
YSO	&	275	&	235	&	0.85$\pm$0.02	&	199	&	0.85$\pm$0.03	&	198	&	0.84$\pm$0.03	&	109	&	0.46$\pm$0.04	&&	275	&	142	&	0.52$\pm$0.03	\\
H{\sc ii} region	&	214	&	167	&	0.78$\pm$0.03	&	129	&	0.77$\pm$0.04	&	149	&	0.89$\pm$0.03	&	126	&	0.75$\pm$0.04	&&	220	&	115	&	0.52$\pm$0.03	\\
H{\sc ii}/YSO	&	71	&	60	&	0.85$\pm$0.04	&	48	&	0.80$\pm$0.06	&	51	&	0.85$\pm$0.05	&	47	&	0.78$\pm$0.06	&&	71	&	40	&	0.56$\pm$0.06	\\
Young/old	&	31	&	17	&	0.55$\pm$0.09	&	10	&	0.59$\pm$0.16	&	8	&	0.47$\pm$0.16	&	4	&	0.24$\pm$0.14	&&	31	&	11	&	0.35$\pm$0.09	\\
\hline
Total &	591	&	479	&	0.81$\pm$0.02	&	387	&	0.81$\pm$0.02	&	406	&	0.85$\pm$0.02	&	286	&	0.60$\pm$0.02	&&	597	&	308	&	0.52$\pm$0.02	\\
\hline\\
\end{tabular}\\
$^{\rm{a}}$ The first, second and third columns give the source type, the number of observations and the total number of sources towards which NH$_3$ emission is detected, respectively. In the fourth column we give the overall detection rate. In columns 5--10 we give the total number of sources where the (1,1) hyperfine structure is clearly detected and the numbers and detection rates for the (2,2) and (3,3) transitions; the detection ratios for these transitions are given as a proportion of the number of NH$_3$ detections (i.e., Column\,3) rather than the number of sources observed. In Columns\,11--13 we give the number of sources observed for H$_2$O masers, the number of detections and the detection rates.
\end{minipage}

\end{center}
\end{table*}
\setlength{\tabcolsep}{6pt}

\subsection{Detection statistics}
\label{sect:det_stats}

Emission from the ammonia (1,1) inversion transition is detected towards 477 sources, with water-maser emission seen towards 308 sources. These correspond to detection rates of 81\,per\,cent and 52\,per\,cent (i.e., 477/591 and 308/597) for NH$_3$ and the H$_2$O masers, respectively. We find that 273 of the water masers are positionally associated with dense gas as traced by the ammonia emission. We present a summary of detection rates 
for the four transitions for the observed sample in Table\,\ref{tbl:detection_stats}, as well as a breakdown into the four source 
classifications. The uncertainties in the detection rates have been calculated using binomial statistics. 

The detection rates for the H{\sc ii}/YSO, H{\sc ii}-region and YSO samples are very similar for the ammonia transitions, with the exception being NH$_3$ (3,3), which is significantly higher ($\sim$60\,per\,cent) for the two H{\sc ii}-region samples compared to the sample consisting exclusively of YSOs. The NH$_3$ (3,3) inversion transition has a higher excitation energy than the  NH$_3$ (1,1) and (2,2) lines ($\sim$125\,K above the ground-state energy), and therefore the higher detection rate for the NH$_3$ (3,3) transition found towards H{\sc ii} regions probably reflects the presence of warmer gas, which is consistent with these regions being more luminous and/or more evolved.

The water-maser detection rates for the YSOs and H{\sc ii}-region samples are also very similar ($\sim$50\,per cent), indicating that the conditions required to produce water-maser emission are equally likely in both phases (cf. \citealt{urquhart2009_h2o}). The detection rate found towards H{\sc ii} regions is comparable with that reported by \citet{kurtz2005} from a set of observations made towards a smaller sample conducted with the 100-m Effelsberg telescope with a similar sensitivity to the data presented here.

The ammonia (1,1) and water-maser detection rates are both significantly lower ($\sim$2 and 3 $\sigma$, respectively) for the young/old classification type,   a strong indication that this group includes a significant number of evolved stars. The H$_2$O maser and NH$_3$ (1,1) and (2,2) detection rates for the young/old group are $\sim$50\,per\,cent lower than for the other three source types, while the detection rate for the NH$_3$ (3,3) transition is a factor of three lower than the H{\sc ii} sub-samples but only a factor of two lower than the YSOs.

\begin{figure}
\begin{center}
\includegraphics[width=0.33\textwidth, trim= 30 0 0 0]{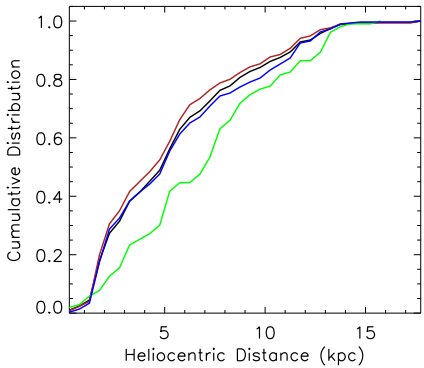}

\caption{\label{fig:distance_CDF} Cumulative distribution functions showing the relative heliocentric distances for the ammonia and water maser detections, and non-detections.The distribution of all NH$_3$ and water maser detections is shown is shown as a thick black line, while the NH$_3$ (1,1) and H$_2$O and non-detections are shown in red, blue and green, respectively.} 

\end{center}
\end{figure}

No ammonia or water maser emission is detected towards $\sim$60 RMS sources classified as either a YSO, H{\sc ii}/YSO or H{\sc ii} region. To investigate these non-detections we present a cumulative distribution plot showing the heliocentric distance distribution of the ammonia and water maser detections and non-detections in Fig.\,\ref{fig:distance_CDF}. Inspection of this plot reveals that the non-detections are preferentially located at larger distances than sources towards which ammonia and/or water masers are detected. For example, we find roughly three-quarters of the ammonia and H$_2$O maser detections are located within 7\,kpc compared with half of the non-detections. Given the small beam filling factor estimated in Sect.\,\ref{sect:filling_factor} and the larger distances associated with the non-detections, we conclude that many of the non-detections result from sources falling below our observational sensitivity. These non-detections are therefore probably due to beam dilution rather than a lack of dense gas associated with any of these sources.

\subsection{Detection rates and luminosity}

\begin{figure*}
\begin{center}
\includegraphics[width=0.33\textwidth, trim= 30 0 0 0]{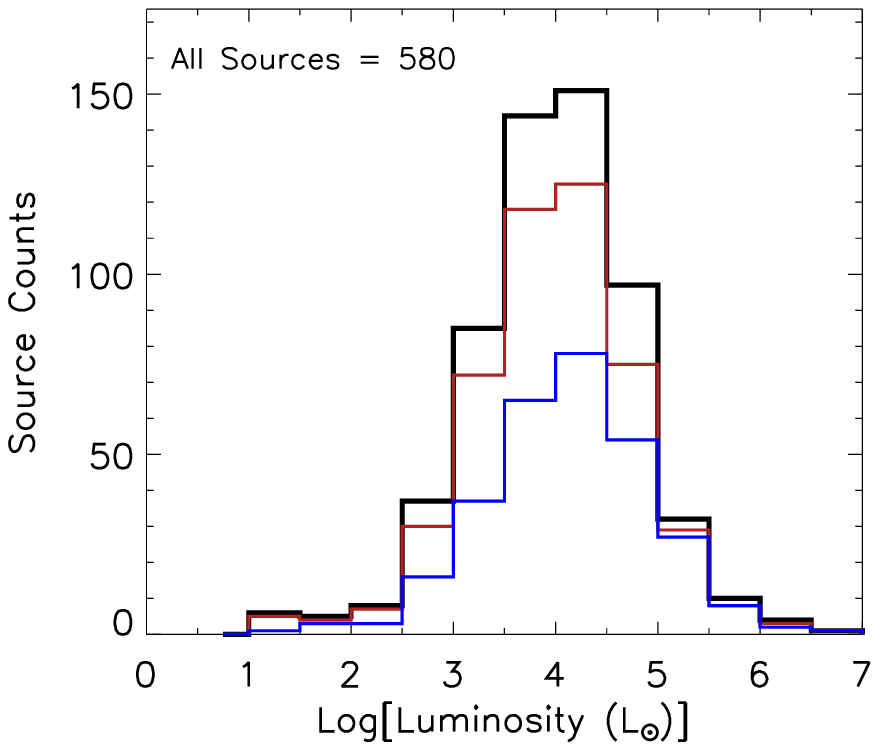}
\includegraphics[width=0.33\textwidth, trim= 30 0 0 0]{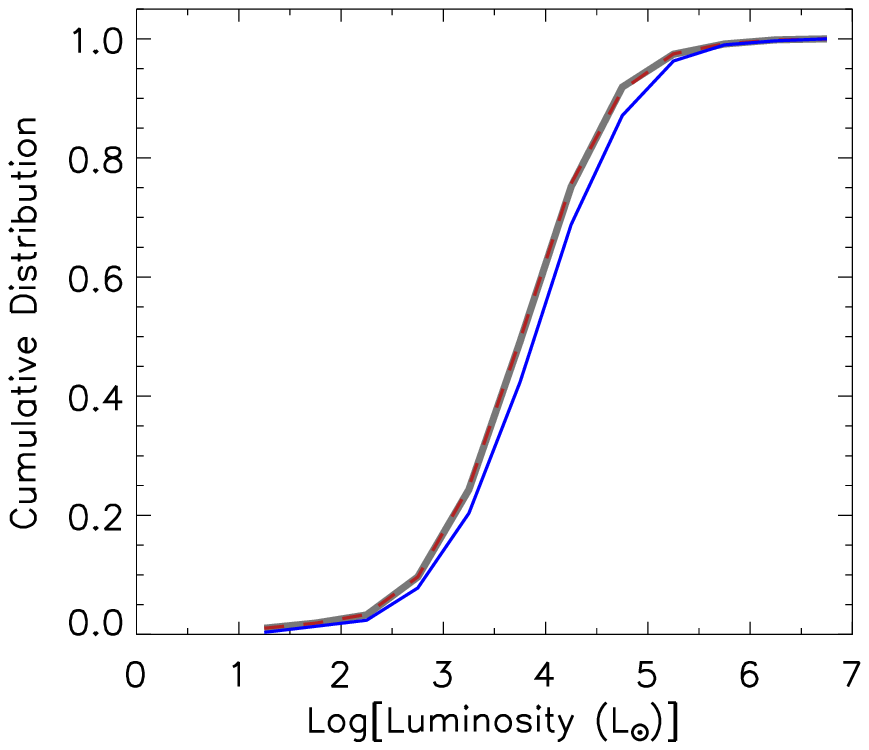}
\includegraphics[width=0.33\textwidth, trim= 30 0 0 0]{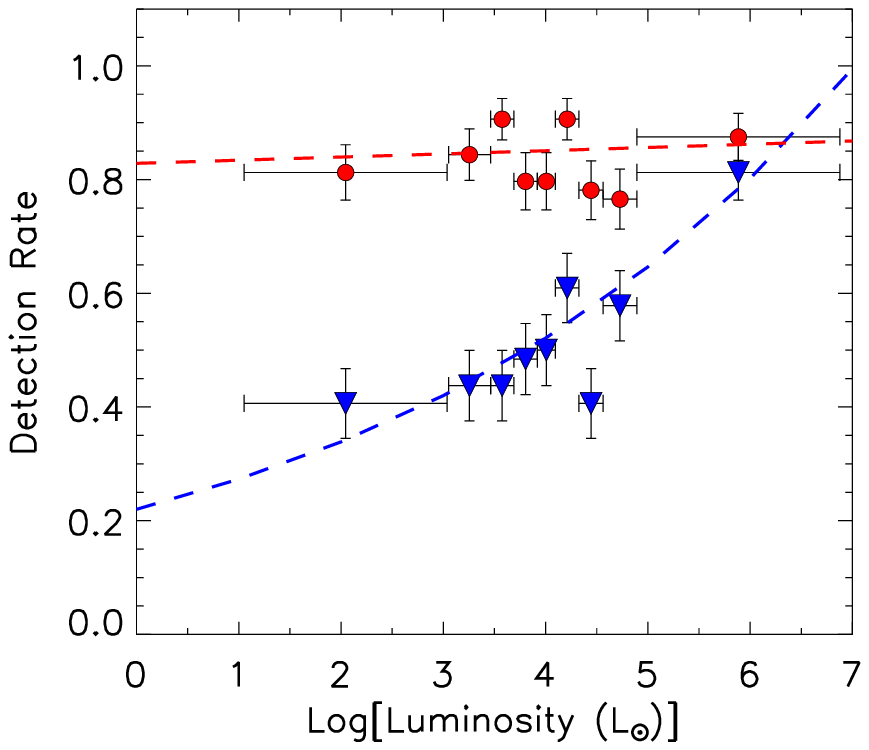}

\caption{\label{fig:luminosity_comparison} Luminosity distribution plots and detection rates as a function of luminosity. In the left and middle panels we present a luminosity histogram and cumulative distribution function for the 580 RMS sources with a derived luminosity (black) and the NH$_3$ (1,1) and H$_2$O detections (coloured red and blue respectively). In the right panel we present the NH$_3$ (1,1) and H$_2$O detection rates as a function of luminosity; the  NH$_3$ (1,1) and H$_2$O data points are shown as red circles and inverted blue triangles respectively. The error bars presented in these plots have been calculated assuming binomial statistics. The red and blue dashed lines show the results of a least-squared fit to the NH$_3$ and H$_2$O data respectively. For the luminosity histogram and cumulative distribution function plot we have used a bin size of 0.5 dex, whilst for the detection statistic plot we use a variable bin width so as to include an even number of sources in each bin.} 

\end{center}
\end{figure*}

In Sect.\,\ref{sect:det_stats} we reported the NH$_3$ (1,1) and H$_2$O detection rates for the observed sample and as a function of each classification type. Overall, we found the NH$_3$ (1,1) and H$_2$O detection rates to be  $\sim$80\,per\,cent and 50\,per\,cent, respectively, and found that these values are similar for both the YSO and H{\sc ii} region samples. In this subsection we will use the RMS source luminosities to investigate whether or not there is any dependence of the detection rates on source luminosity. 

In the left and middle panels of Fig.\,\ref{fig:luminosity_comparison} we present the luminosity distribution for the 580 RMS sources with a derived luminosity (thick black line; hereafter we will refer to this sample as the RMS luminosity sample) and of the sources towards which NH$_3$ (red line) and H$_2$O masers (blue line) are detected. In the right panel of this figure we present the NH$_3$ (red circles) and water maser (blue triangles) detection rates as a function of luminosity. 

Comparing the luminosity distribution RMS luminosity sample with that of the NH$_3$-detected sample reveals them to be indistinguishable from each other. However, comparing the luminosity distribution RMS luminosity sample with that of the H$_2$O-detected sample we notice a slight increase in the number of more luminous sources associated with H$_2$O masers. A Komolgorov-Smirnov (KS) test was used to compare the H$_2$O masers sample with the RMS luminosity sample but found them not to be measurably different.

The plot presented in the right panel of Fig.\,\ref{fig:luminosity_comparison} reveals the NH$_3$ (1,1) detection rate to be constant over the whole range of luminosities. However, this plot implies that the H$_2$O detection rate has a positive dependence on the source luminosity, with the detection rate increasing with increasing luminosity.  The correlation coefficient is $\sim$0.8 which, for $N=9$, is significant with a probability of $\ll 1$\,per\,cent of arising by chance. A linear least-squares fit to the data in Fig.\,10 predicts the following relationship:

\begin{equation}
{\rm{Log}}\left[\frac{{\rm{RMS}}_{\rm{{H_2O}}}}{{\rm{RMS_{All}}}}\right] = (0.094\pm0.014)\times {\rm{log}}(L_{\rm{bol}}) - (0.66\pm0.07) 
\end{equation}

\noindent where $L_{\rm{bol}}$ is the bolometric luminosity in solar luminosities and RMS$_{\rm{H_2O}}$/RMS$_{\rm{All}}$ is the expected detection rate as a function of luminosity with a sensitivity of $\sim$0.1\,Jy beam$^{-1}$ (1$\sigma$). The resultant fit to the H$_2$O maser detection rate is shown as a dashed blue line in the right panel of Fig.\,\ref{fig:luminosity_comparison}. The detections rates of the YSO and H{\sc ii} region samples indicate similar trends, however, lower sample numbers and the larger associated uncertainties prevent us from drawing any firm conclusions for these samples.

\subsection{Derived NH$_3$ parameters}
\label{sect:nh3_parameters}

\setlength{\tabcolsep}{4pt}

\begin{table*}

\begin{center}\caption{Mean values derived from the NH$_3$ transitions. The errors shown in 
parenthesis are the standard error on the mean.}
\label{tbl:parameter_stats}
\begin{minipage}{\linewidth}
\begin{tabular}{lccccccccccc}
\hline \hline
 & T$_{\rm{rot}}$&  T$_{\rm{kin}}$& $\tau$ & Log(N(NH$_3$)) & $T_{\rm{mb(1,1)}}$  &$\Delta{\rm{V}}_{\rm{(1,1)}}$ & $T_{\rm{mb(2,2)}}$  & $\Delta{\rm{V}}_{\rm{(2,2)}}$ & $T_{\rm{mb(3,3)}}$  & $\Delta{\rm{V}}_{\rm{(3,3)}}$ \\
Source Type & (K)& (K)& & (cm\,s$^{-2}$ & (K) &(\kms) &(K) & (\kms) &(K) & (\kms) \\
 
\hline 
IRDC cores$^a$	&	13.34 (0.54)	 &	13.76 (0.56)	 &	4.25 (0.92)	 &	15.43 (0.16)	 &	3.12 (0.59)	 &	1.45 (0.14)	 &	1.14 (0.41)	 &	1.51 (0.23)	 &	1.42 (0.39)	 &	$\cdots$	 \\
YSO	&	17.84 (0.24)	 &	20.12 (0.35)	 &	2.25 (0.11)	 &	15.22 (0.02)	 &	1.38 (0.07)	 &	1.66 (0.05)	 &	0.66 (0.04)	 &	1.70 (0.10)	 &	0.48 (0.04)	 &	3.13 (0.09)	 \\
H{\sc ii} region	&	20.89 (0.35)	 &	24.58 (0.57)	 &	2.33 (0.14)	 &	15.44 (0.03)	 &	1.16 (0.06)	 &	2.20 (0.11)	 &	0.64 (0.04)	 &	2.37 (0.17)	 &	0.50 (0.04)	 &	4.21 (0.44)	 \\
H{\sc ii}/YSO	&	20.89 (0.43)	 &	24.41 (0.69)	 &	1.81 (0.19)	 &	15.24 (0.05)	 &	1.11 (0.11)	 &	2.41 (0.21)	 &	0.64 (0.08)	 &	2.20 (0.22)	 &	0.40 (0.07)	 &	3.28 (0.18)	 \\
Young/old	&	15.74 (0.59)	 &	17.18 (0.75)	 &	2.91 (0.61)	 &	15.29 (0.12)	 &	1.08 (0.22)	 &	0.74 (0.11)	 &	0.52 (0.14)	 &	1.41 (0.19)	 &	0.31 (0.13)	 &	2.47 (0.47)	 \\

\hline

 	&	19.22 (0.20)	 &	22.12 (0.30)	 &	2.24 (0.08)	 &	15.30 (0.02)	 &	1.27 (0.04)	 &	1.91 (0.06)	 &	0.65 (0.03)	 &	2.00 (0.08)	 &	0.47 (0.03)	 &	3.62 (0.20)	 \\

\hline

\end{tabular}\\
$^{\rm{a}}$ The values presented in this row have been taken from \citet{pillai2006}. 

\end{minipage}

\end{center}
\end{table*}
\setlength{\tabcolsep}{6pt}

In Table\,\ref{tbl:parameter_stats} we present mean values for the parameters derived in Sect.\,\ref{sect:nh3_properties} for the whole observed sample and broken down by source classification. In this table we include the mean values for the `young/old' sample for completeness and, since the values of the various parameters associated with the H{\sc ii}/YSO sample are almost indistinguishable from those derived for the H{\sc ii} region sample, we will not discuss the properties of either of these two samples in detail. We complement the results presented in Table\,\ref{tbl:parameter_stats} with values reported by \citet{pillai2006} for a sample of IRDC cores that are \emph{not} coincident with strong compact mid-infrared emission and have masses of a few hundred \msun. These IRDC cores therefore constitute a sample of massive starless cores (as indicated by the lack of mid-infrared emission that would indicate the presence of a YSO or H{\sc ii} region) that pre-date the YSO and H{\sc ii} region phases. Combined, these three well-defined classes of objects (IRDC cores, YSOs, and H{\sc ii} regions) should broadly cover the main evolutionary stages of massive star formation.

\begin{figure}
\begin{center}
\includegraphics[width=0.23\textwidth, trim= 30 0 0 0]{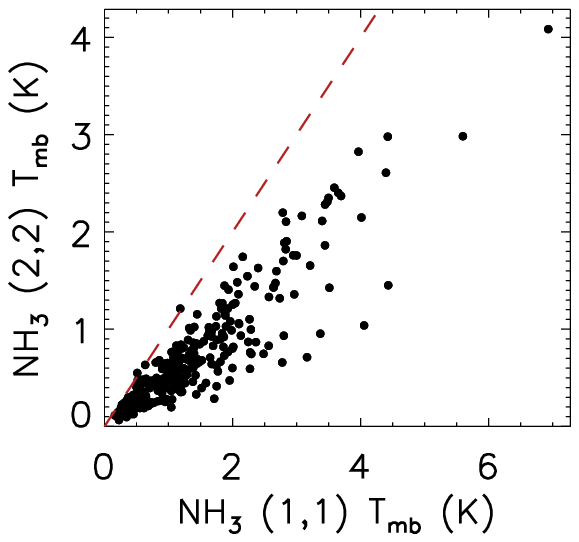}
\includegraphics[width=0.23\textwidth, trim= 30 0 0 0]{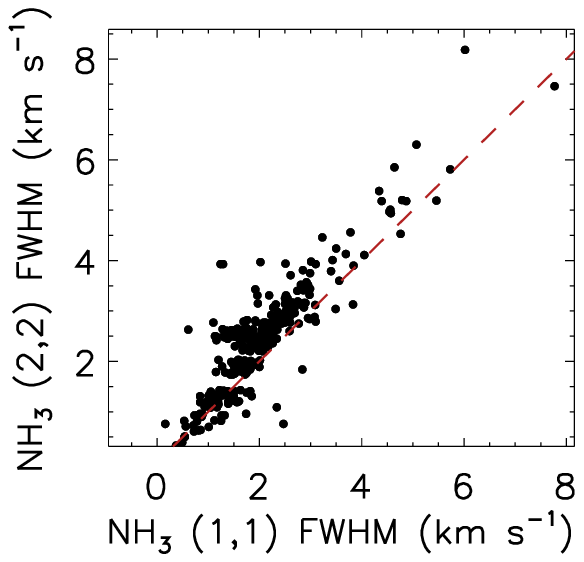}
\includegraphics[width=0.23\textwidth, trim= 30 0 0 0]{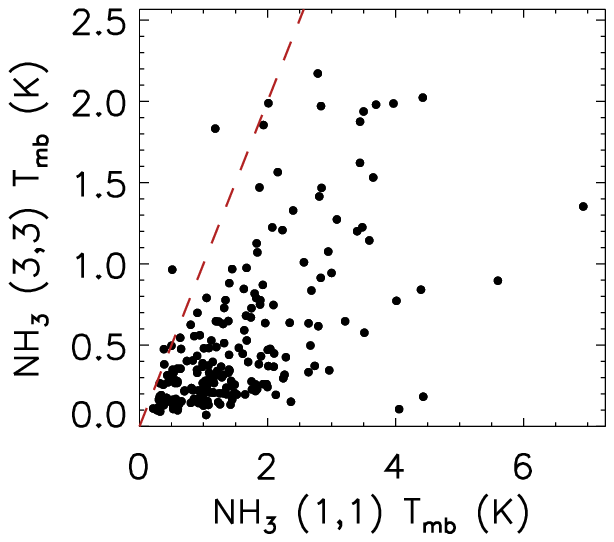}
\includegraphics[width=0.23\textwidth, trim= 30 0 0 0]{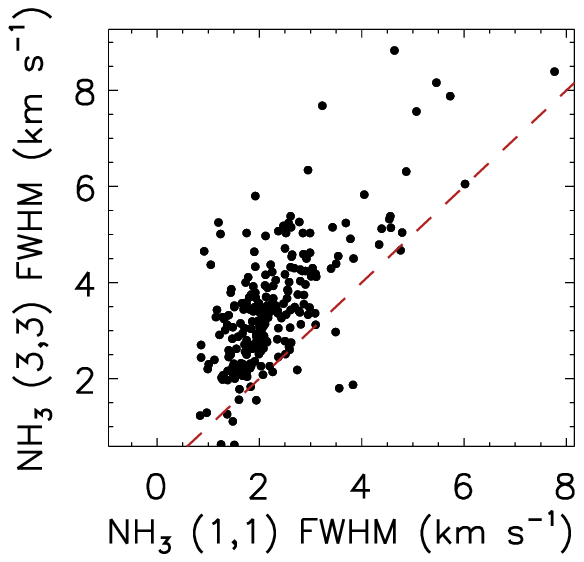}

\caption{\label{fig:tmb_fwhm_comp} Comparison plots of the main beam temperatures and FWHM line widths of the NH$_3$ (1,1), (2,2) and (3,3) transitions. The dashed red lines mark the line of equality.} 

\end{center}
\end{figure}

Looking at the global properties of the whole observed sample we find that the mean kinetic temperature is $\sim$20\,K, which is twice what would be expected from heating from the interstellar radiation field alone ($\sim$10\,K; \citealt{evans1999}). These temperatures would require the presence of either an internal or external heating source, and since we know that the observed objects are associated with embedded massive stars, the former is considered more likely. We also find the mean line width for the NH$_3$ (1,1) detections is $\sim$2\,\kms, which is approximately ten times broader than the thermal linewidth (approximately 0.2\,\kms\ for gas temperatures $\sim$20\,K). These line widths must be due to systematic motions within the gas such as infall, molecular outflows, stellar winds, cloud rotation or internal random motions referred to as \emph{turbulence}.    

Fig.\,\ref{fig:tmb_fwhm_comp} presents scatter plots comparing the $T_{\rm{mb}}$ and line widths of the \nhthree\ (1,1), (2,2) and (3,3) transitions, revealing a strong correlation between the (1,1) and (2,2) line widths and $T_{\rm{mb}}$. Importantly, the line width ratio for the  (1,1) and (2,2) transitions is  close to unity indicating that the two transitions are tracing a similar volume of gas --- this was implicitly assumed when deriving the physical parameters in Sect.\,2.2.1. Comparison of the NH$_3$ (1,1)  and (3,3) transition data reveals a much lower level of correlation for both parameters, with the NH$_3$ (3,3) transition showing systematically lower main beam temperatures and broader line widths. The higher gas temperatures required to excite the \nhthree\ (3,3) transitions would suggest that the observed emission is emitted from a smaller volume of warmer gas located nearer to the embedded object than the \nhthree\ (1,1) emission, which is tracing the cooler outer envelope. The broader line widths found for the NH$_3$ (3,3) emission are also consistent with this transition tracing warmer and/or more turbulent gas closer to the embedded object.

Inspection of the mean parameters determined for the IRDC cores, YSOs, and H{\sc ii} regions reveals trends for increased linewidth and kinetic temperature with evolution phase. Similar trends have been reported in the literature and these could be related to the evolutionary phase of the embedded object as has been suggested by \citet{sridharan2005} from a sample of 56 candidate high-mass starless cores, high-mass protostellar objects and UC\,H{\sc ii} regions. However, as we will show in the following section this interpretation is overly simplistic.

\subsection{NH$_3$ correlations}

In the previous subsection we noted a number of trends seen in the derived parameters of the IRDC cores, YSO and H{\sc ii} regions samples. We observed a trend for increasing kinetic temperature and line width as we moved between the IRDC core, YSO,  and H{\sc ii} region samples. Since these three classifications represent an evolutionary sequence, these trends would seem to suggest an underlying connection. However, \citet{mottram2011a}  showed that there is a difference in the luminosity distributions of the YSO and H{\sc ii} region samples, with the H{\sc ii} regions typically being an order of magnitude more luminous. Given the luminosity difference between the samples it is possible that the trends seen in the derived parameters are actually related to the luminosity of the samples rather being associated with evolutionary progression.

\begin{figure*}
\begin{center}
\includegraphics[width=0.33\textwidth, trim= 30 0 0 0]{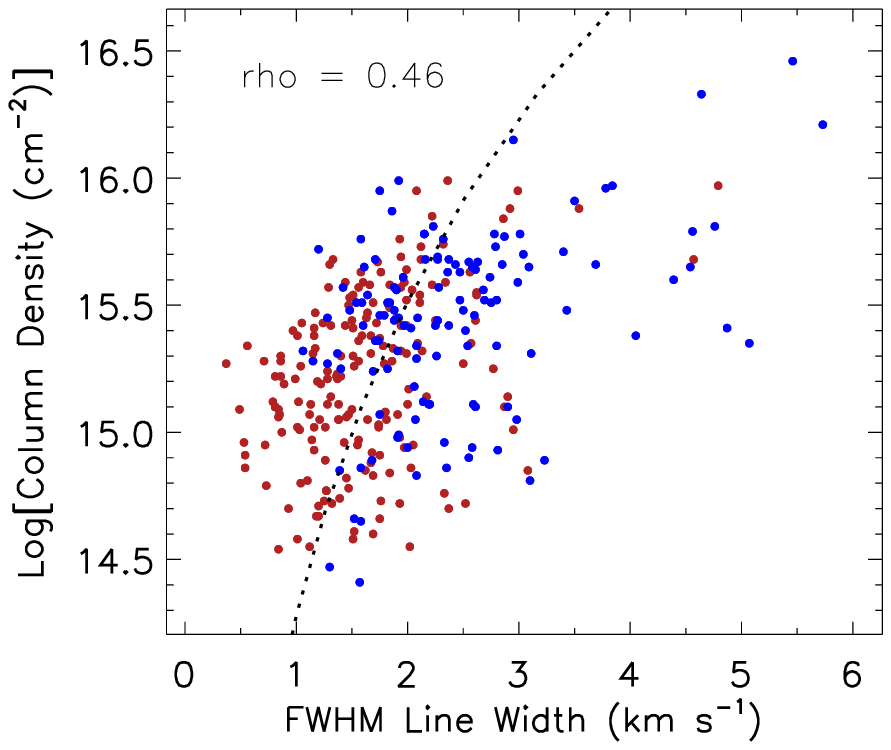}
\includegraphics[width=0.33\textwidth, trim= 30 0 0 0]{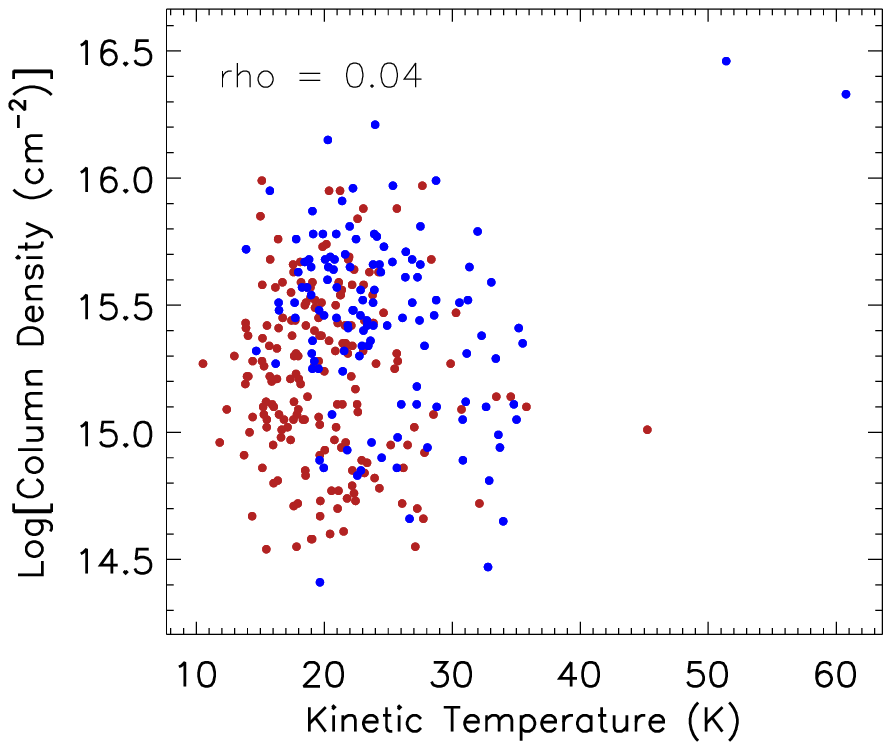}
\includegraphics[width=0.33\textwidth, trim= 30 0 0 0]{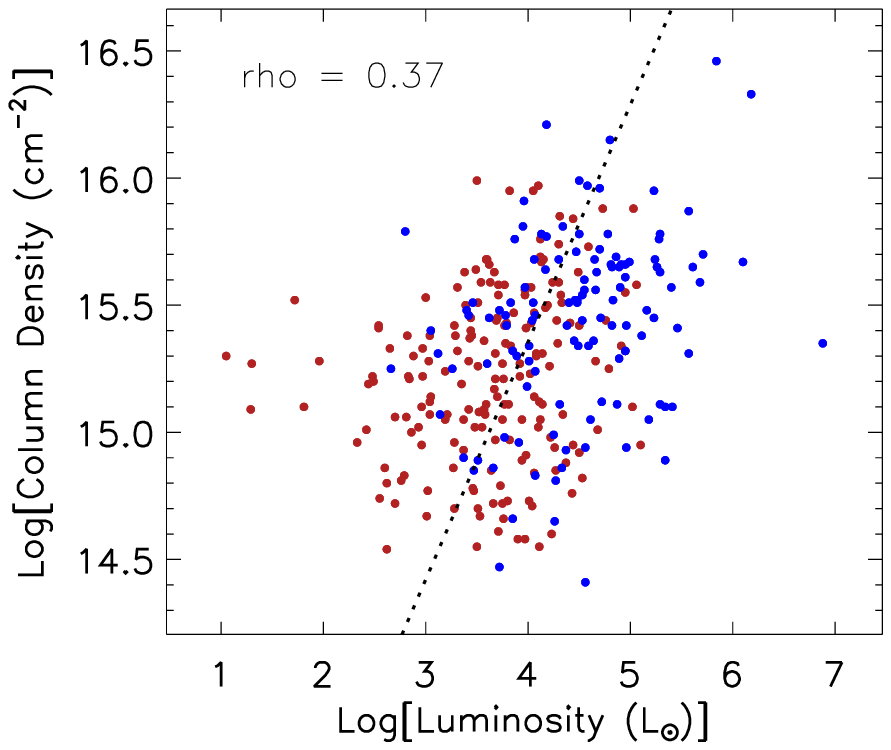}

\includegraphics[width=0.33\textwidth, trim= 30 0 0 0]{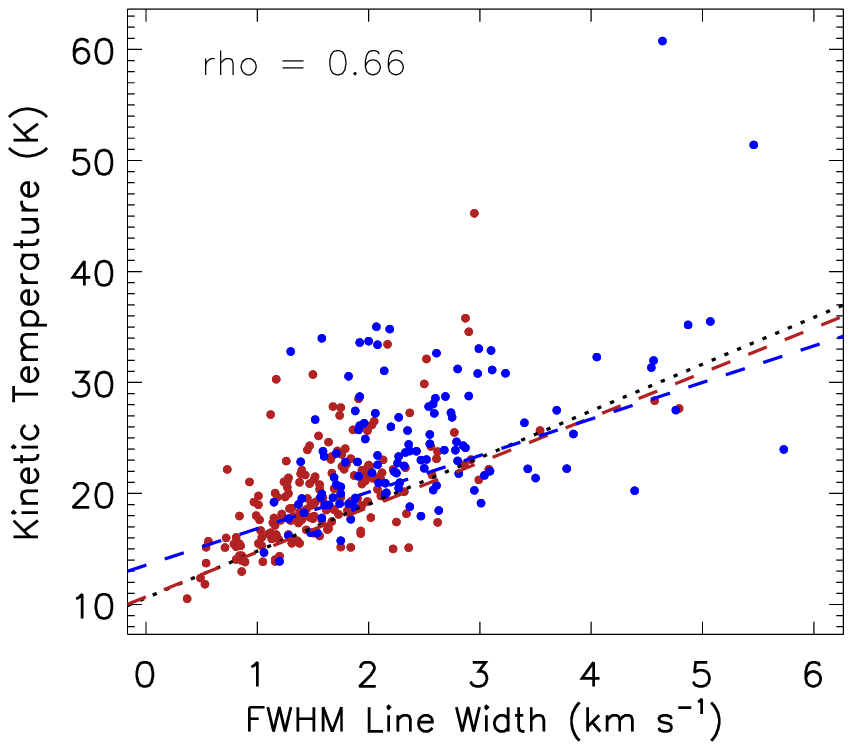}
\includegraphics[width=0.33\textwidth, trim= 30 0 0 0]{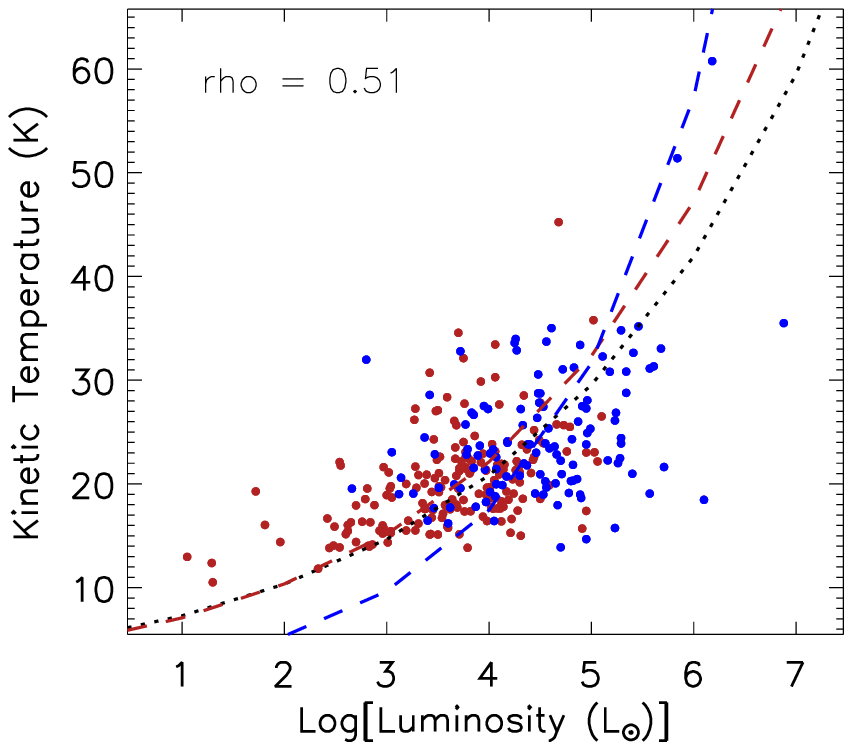}
\includegraphics[width=0.33\textwidth, trim= 30 0 0 0]{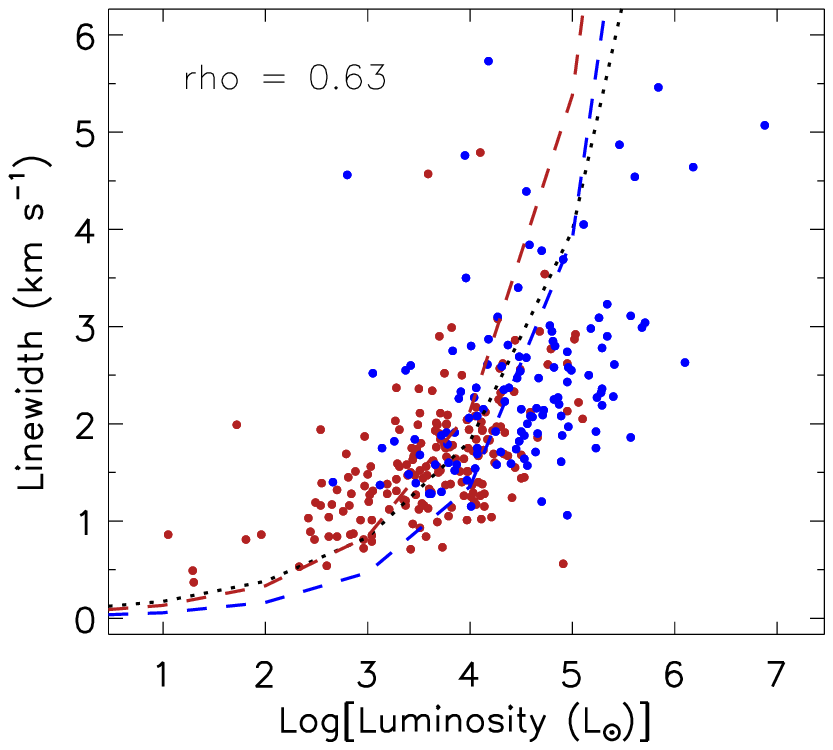}

\caption{\label{fig:temp_linewidth_ratio} Scatter plots showing the correlation between the column density, $T_{\rm{kin}}$, $\Delta{\rm{V}}$ and $L_{\rm{bol}}$. The YSO and H{\sc ii} region sub-samples are shown in red and blue respectively. The Spearman correlation coefficients are given in the top left corner of each plot. The black dotted line shows the result of linear least-squared fits to all of the plotted data, whilst the red and blue dashed lines show the results of linear least-squared fits to the YSO and H{\sc ii} region sub-samples respectively.} 

\end{center}
\end{figure*}

\begin{table}
\begin{center}\caption{Gradients and intercepts of linear least-squared best fits to the plots presented in Fig.\,\ref{fig:temp_linewidth_ratio}. These fits have been performed in log-log space.  }
\label{tbl:correlation_stats}
\begin{minipage}{\linewidth}
\begin{tabular}{cccccc}
\hline \hline
Plot & Group  &     Intercept  &  Gradient       &     rho     \\
 
\hline
N(NH$_3$) vs. $\Delta{\rm{V}}$  &    All  &	14.28$\pm$0.31	&	4.10$\pm$1.16	&	0.46 \\
\hline
N(NH$_3$) vs. $L_{\rm{bol}}$     &    All  &	11.62$\pm$0.25	&	0.93$\pm$0.06	&	0.37	\\

\hline

 									&	All	&	0.71$\pm$0.03	&	0.15$\pm$0.01	&	0.51	\\
$T_{\rm{kin}}$ vs $L_{\rm{bol}}$	&	YSO	&	0.69$\pm$0.04	&	0.16$\pm$0.01	&	0.49	\\
									&	H{\sc ii}	&	0.21$\pm$0.15	&	0.26$\pm$0.03	&	0.30	\\

\hline
									&	All	&	10.59$\pm$0.07	&	4.21$\pm$0.05	&	0.66	\\
$T_{\rm{kin}}$ vs. $\Delta{\rm{V}}$		&	YSO	&	10.69$\pm$0.07	&	4.04$\pm$0.05	&	0.63	\\
									&	H{\sc ii}	&	13.55$\pm$0.32	&	3.29$\pm$0.15	&	0.49	\\

\hline
									&	All	&	-1.10$\pm$0.06	&	0.34$\pm$0.02	&	0.63	\\
$\Delta{\rm{V}}$ vs. $L_{\rm{bol}}$ 		&	YSO	&	-1.28$\pm$0.11	&	0.40$\pm$0.03	&	0.56	\\
									&	H{\sc ii}	&	-1.71$\pm$0.25	&	0.46$\pm$0.06	&	0.44	\\
    
\hline
\end{tabular}\\
\end{minipage}

\end{center}
\end{table}

To investigate the correlation between the derived NH$_3$ parameters and source type and/or luminosity, we present six scatter plots in Fig.\,\ref{fig:temp_linewidth_ratio}. In the upper panels of this figure we compare the sources' column densities with their $T_{\rm{kin}}$, $\Delta{\rm{V}}$ and $L_{\rm{bol}}$, whilst in the lower panels we compare the latter three parameters with each other. In these plots we show the distribution of the YSO and H{\sc ii} region sub-samples in red and blue, respectively, and give the Spearman correlation coefficients (rho) in the top left corner of each plot; all non-zero correlation coefficients are significant with a probability of arising by chance of $\ll 1$\,per\,cent. We find a correlation between five of the six pairs of plotted parameters and in these cases we have performed a linear fit to the combined YSO and H{\sc ii} region samples which is shown as dotted black line. In cases where we find a moderate correlation (i.e., 0.5$<$rho $<$0.75) we have made separate linear fits to the YSO and H{\sc ii} region sub-samples to investigate possible differences in the overall trends; these are shown as dashed red and blue lines respectively. The intercepts and gradients for all of these fits are presented in Table\,\ref{tbl:correlation_stats}. 

Visual inspection of the upper panels of Fig.\,\ref{fig:temp_linewidth_ratio} reveal there is no correlation between column density and $T_{\rm{kin}}$ and only a weak correlation between column density and the other two parameters. The correlation between the $T_{\rm{kin}}$, $\Delta{\rm{V}}$ and $L_{\rm{bol}}$ parameters is significantly stronger than found for the column density with values of rho $>$ 0.5. However, we note that the correlation for these parameters is lower for the YSO and H{\sc ii} region samples with only a weak correlation being found for the H{\sc ii} regions. The correlation between the $T_{\rm{kin}}$ and $\Delta{\rm{V}}$ (lower left panel of Fig.\,\ref{fig:temp_linewidth_ratio}) is particularly strong with a correlation coefficient of 0.66. Moreover, the YSOs and H{\sc ii} regions dominate either end of what appears to be fairly continuous distribution. This would seem to support the trend of increasing line width and kinetic temperature with evolution touched upon in Section\,\ref{sect:nh3_parameters}. However, turning our attention to the lower middle and right panels of Fig.\,\ref{fig:temp_linewidth_ratio} we find that both of these parameters are also correlated with the sources' bolometric luminosity.

\citet{churchwell1990} found a similar, although weaker, correlation between the kinetic temperature and linewidth with far-infrared luminosity from NH$_3$ observations of a smaller sample of UC\,H{\sc ii} regions (84) made with the Effelsberg 100-m telescope. The authors suggested that the line width-luminosity correlation can be understood as either: 1) the amount of mechanical energy deposited into the ambient molecular cloud (e.g., via molecular outflows, stellar winds) being directly proportional to the luminosity of the embedded source; or 2) the line width-clump mass relationship (i.e., \citealt{larson1981}), with the more luminous, and presumably more massive, stars forming out of more massive clumps.  The correlation between kinetic temperature and luminosity would indicate that the heating of the gas is dominated by the radiation deposited by the embedded star into its natal environment.

The distributions presented in the right panels of Fig.\,\ref{fig:temp_linewidth_ratio} suggest that there is a dependence of line width and kinetic temperature on bolometric luminosity. It is clear from these correlation plots that the embedded sources are having a measurable effect on their surrounding molecular environment. Given that these observations are sampling clumps with sizes ranging from 0.1-0.5\,pc it is reasonable to conclude that it is the energy output of a young massive star that is driving the increases in temperature and line width. However, it is not yet clear if there is also a weaker underlying dependence on the evolutionary phase of the embedded sources.

\begin{figure}
\begin{center}
\includegraphics[width=0.23\textwidth, trim= 30 0 0 0]{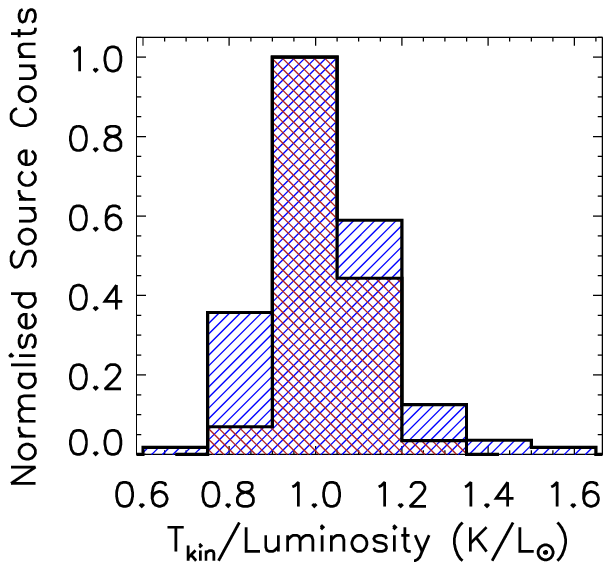}
\includegraphics[width=0.23\textwidth, trim= 30 0 0 0]{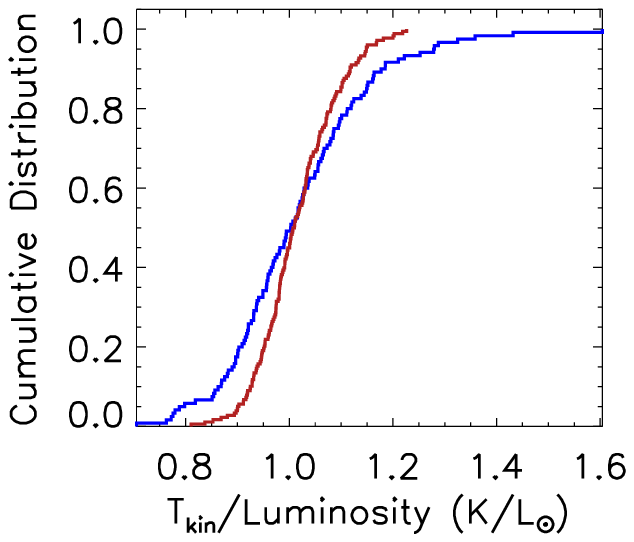}
\includegraphics[width=0.23\textwidth, trim= 30 0 0 0]{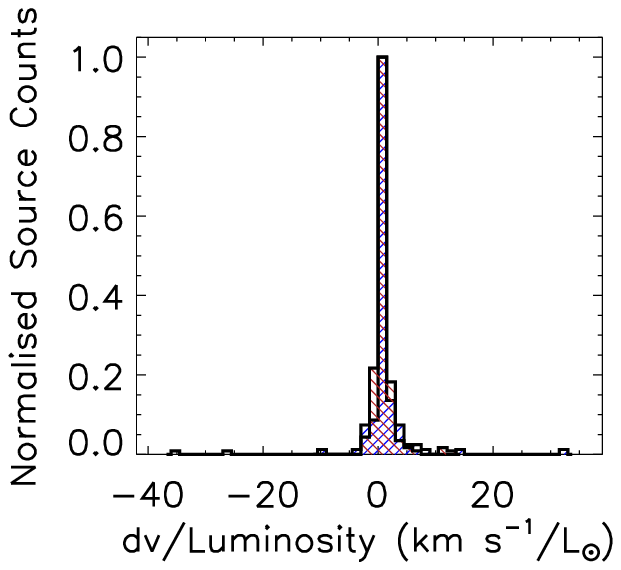}
\includegraphics[width=0.23\textwidth, trim= 30 0 0 0]{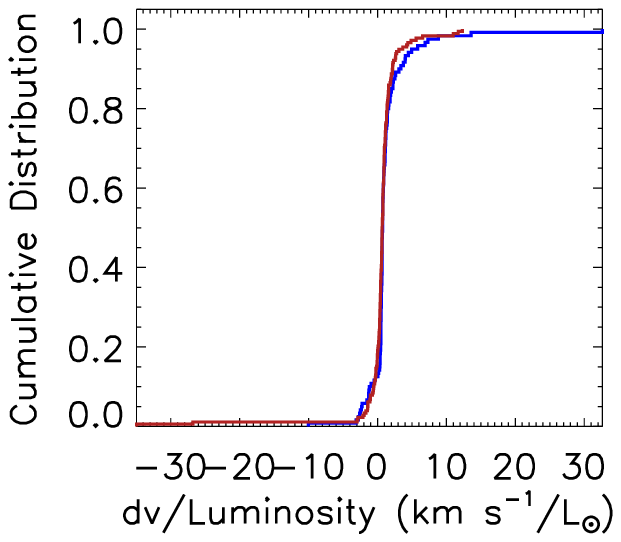}

\caption{\label{fig:dvlsr_kin_ks} Histograms and cumulative distribution plots showing the differences between the YSO and H{\sc ii} region samples after the luminosity dependence has been removed. The distributions of the YSO and H{\sc ii} regions samples are shown in red and blue respectively.} 

\end{center}
\end{figure}

Inspection of the plots and fit parameters presented in Table\,\ref{tbl:correlation_stats} would suggest there is a significant difference between the gas temperature and line width, and the gas temperature and the luminosity for the H{\sc ii} region and YSO sub-samples. There is no significant difference between the  line width-luminosity relationship for the two samples, which would suggest that it is differences in the kinematic gas temperature that is the parameter of interest. In an attempt to investigate the gas temperature and line width differences for the H{\sc ii} regions and YSO sub-samples we have removed the luminosity dependence using their respective fit parameters; this effectively normalises both parameters. In Fig.\,\ref{fig:dvlsr_kin_ks} we present normalised histograms and cumulative distribution plots showing the differences between the YSO and H{\sc ii} region samples after the luminosity dependence has been removed. These plots reveal little difference between the two normalised parameters. Komolgorov-Smirnov (KS) tests used to compare the two samples found that the probabilities that YSOs and H{\sc ii} regions were drawn from the same parent population were 3\,per\,cent and 4\,per\,cent for the normalised kinetic temperature and line width parameters respectively. We therefore conclude that the observed trends in Table 7 are more likely to be due to the energy output of the central source and/or the line width-clump mass relationship than the evolutionary state of the embedded source.

\subsection{H$_2$O maser correlations}
\label{sect:h2o_correlations}

Given the ubiquitous nature of water masers within the Galactic plane it is possible that a number of the detected masers are the result of chance line of sight alignments within the field or a bright maser located within a telescope sidelobe rather than genuine association; however, given the GBTs off-axis design, sidelobe contamination is significantly lower than for traditional single-dish telescopes and so the latter possibility is less of a concern\footnote{\label{gbtfootnote}`The Proposer's Guide for the Green Bank Telescope'.}. In this section we will investigate the degree of correlation between the water maser and NH$_3$ emission in order to ascertain whether they are arising from the same molecular gas, and if so, whether there is a direct connection between the maser and the embedded massive young star.

\begin{figure}
\begin{center}
\includegraphics[width=0.45\textwidth, trim= 30 0 0 0]{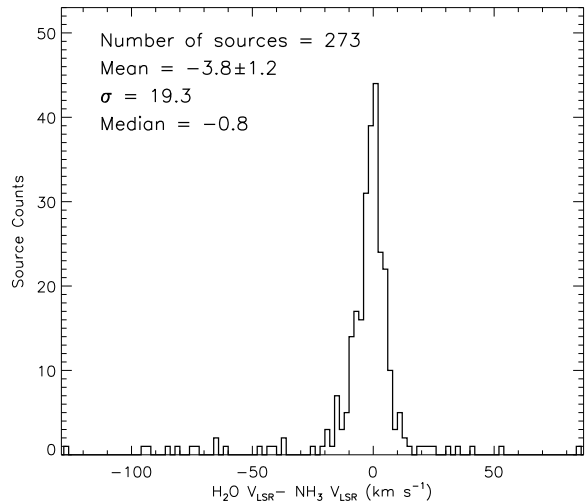}

\caption{\label{fig:nh3_h2o_hist} Distribution of the differences between the velocity of the brightest water maser spot and that of the ammonia emission.} 

\end{center}
\end{figure}

In Fig.\,\ref{fig:nh3_h2o_hist} we present a histogram of the difference in velocity between the most intense water maser spot and that of the dense molecular gas traced by the NH$_3$ emission. This plot reveals that the velocities of the two transitions are in excellent agreement, with a Pearson correlation coefficient of 0.92. The mean velocity difference is --3.8 with a standard deviation of $\sim$20\,\kms. Computing the standard error on the mean, we find that this small, negative offset from zero is significant at the $>$ 3-$\sigma$ level.  However, the standard deviation is inflated by a small number of high-velocity outliers in the distribution.  If we exclude the 25 sources whose the velocity difference is larger than 20\,\kms, we find that the mean and standard deviation reduce to $\sim-$1.1\,\kms\ and 6\,\kms, respectively. The associated standard error on the mean is $\sim$ 0.38\,\kms, so the revised mean is still nearly 3 $\sigma$ away from zero.  Thus there appears to be a real offset from zero in the mean relative velocity of the maser emission. The distribution also has a slight skew ($\sim$0.26\,\kms) towards blue-shifted sources, with a negative median value ($-$0.82\,\kms).  So not only are there more blue-shifted than red-shifted masers but they also have higher relative velocities.  This probably means that the blue-shifted maser spots tend to be slightly brighter than their red-shifted counterparts and so more of them are detected.

\begin{figure}
\begin{center}
\includegraphics[width=0.45\textwidth, trim= 30 0 0 0]{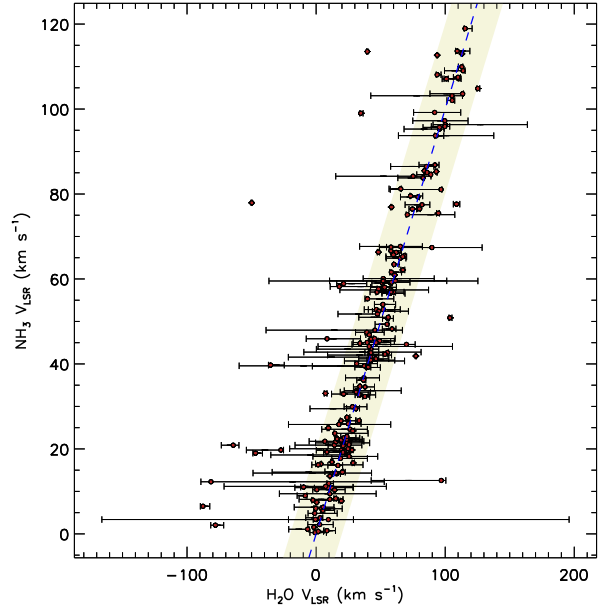}

\caption{\label{fig:nh3_h2o_plot} Comparison of the velocities of the NH$_3$ emission and H$_2$O maser. The red circles show the velocity of the NH$_3$ emission and the velocity of the brightest H$_2$O maser spot, whilst the horizontal black line indicates the velocity range over which the maser emission is detected. The dashed line shows the line of equality where the two velocities are equal and the the light yellow stripe running through the plot indicates region where the H$_2$O maser velocity is within $\pm$20\,\kms\ of the NH$_3$ emission.} 

\end{center}
\end{figure}

In Fig.\,\ref{fig:nh3_h2o_plot} we plot the NH$_3$ velocities against the velocity of the brightest maser spot. The black horizontal lines shown in this plot indicate the velocity range over which maser spots are seen towards each source. The data presented in Fig.\,\ref{fig:nh3_h2o_plot} illustrate the correlation between the molecular gas and the range of maser velocities and the excellent correlation between the velocity of the brightest maser spots and the molecular gas. For the vast majority of sources, the peak maser velocity and NH$_3$ velocity are within 20\,\kms\  of each other, but there are 25 sources in which the velocity difference is larger.  The light yellow stripe running through the plot presented in  Fig.\,\ref{fig:nh3_h2o_plot} indicates the region where the H$_2$O maser velocity is within $\pm$20\,\kms\ of the NH$_3$ emission. This number is further reduced if we take the total velocity range of the maser spots seen towards each source into account; of these 25 sources we find that the NH$_3$ velocity of five lie within the maser velocity range, and five others are located within 20\,\kms\ of a maser spot.

There are 15 sources ($\sim$6\,per\,cent of our sample) for which there is $>$20\,\kms\ difference between the velocity of the nearest maser spot and the velocity of the molecular cloud. Targeted NH$_3$ and H$_2$O observations conducted by \citet{churchwell1990} and \citet{anglada1996} (that have approximately the same resolution and sensitivity to the observations we present here) reported the detection of a similar proportion of sources with large velocity offsets.  \citet{churchwell1990} concluded that these are simply serendipitously matched sources that are located nearby on the sky but which arise from physically distinct regions located along the same line of sight. However, the strong variability of water masers and/or limited sensitivity could also explain the large velocity differences.

In a recent study of water masers by \citet{breen2010}, large changes in the velocity of the brightest maser peak were found for four sources by comparing the emission spectra taken in two epochs separated by a period of 10 months. Taking source G336.983$-$0.183 as an example, they found only a single maser spot at approximately $-$75\,\kms\ in their 2003 observations. However, in the spectrum obtained in 2004 they found two maser spots at approximately $-$75 and 45\,\kms. It is therefore possible that the strong variability of water masers can sometimes result in the large velocity differences observed towards some sources. So although we find large velocity differences between a small number of water masers and the molecular gas it is not yet clear that these arise from physically separate regions. It is clear, however, that there is a strong correlation between velocities of the NH$_3$ and H$_2$O transitions for $\sim$95\,per\,cent of the sources and it is therefore likely that they are associated with each other.

By making the assumption that all of the detected water masers are associated with the molecular clouds, we can investigate the distribution of high-velocity features, which we define as those which vary from the clouds' systemic velocity by more than 30\,\kms\ (c.f. \citealt{caswell2010a}). We find 10 sources that show both high-velocity red- and blue-shifted features, 33 that show only blue-shifted features, 17 that show only red-shifted features and 213 sources that show no high-velocity features. We note that there are almost twice as many sources with high-velocity blue-shifted features. A similar distribution of high-velocity red- and blue-shifted features has been reported by \citet{caswell2010a} who interpreted the excess of blue-shifted features as a sign of the youth of the embedded object as there was also a strong correlation with these blue-shifted features and methanol masers, which are considered to be exclusively associated with the earliest phases of high-mass star formation.

This excess in blue-shifted, high-velocity maser emission reflects the small but significant skew in the general distribution described above and is likely to be related to expansion or contraction in the core or envelope gas.  Further analysis would require detailed modelling of the velocity field and density distribution.  We compared the maser parameters with the column densities and line widths obtained from the ammonia data and as a function of source type to identify any correlations that might provide any further insight into this excess of blue-shifted emission, however, these proved inconclusive.

\begin{figure}
\begin{center}
\includegraphics[width=0.90\linewidth, trim= 30 0 0 0]{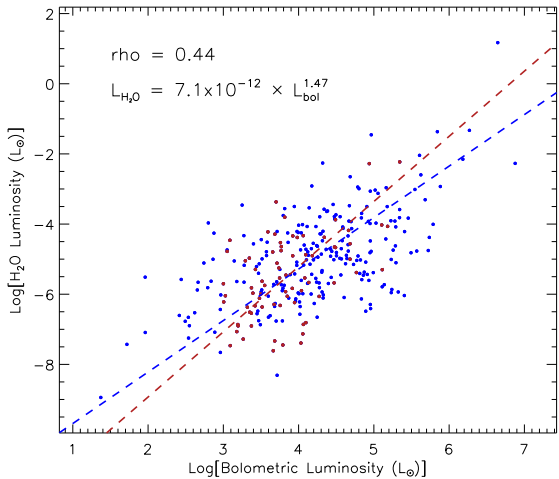}

\caption{\label{fig:h2o_luminosity_comparison} Comparison of the bolometric luminosity of RMS sources with the isotropic water maser luminosity. The blue dots show the luminosities of all maser detections for which we have resolved the distance ambiguity while the red dots show the source luminosities of a distance limited sample (see text for details). The dashed line shows the linear bisector least-squares fit to the data. The partial Spearman correlation coefficient and the and results of the fit to the data are given in the top left of the plot.} 

\end{center}
\end{figure}

Having established that the observed water-maser emission is very likely to be associated with the same molecular clouds within which the MYSOs and H{\sc ii} regions are forming, the next step is to see if we can identify a direct connection between the masers and the embedded young massive star. In Fig.\,\ref{fig:h2o_luminosity_comparison} we present a scatter plot comparing the isotropic maser luminosity calculated in Section\,\ref{sect:Water_maser_properties} with the RMS luminosity. On this plot we show the distribution of all of the H$_2$O detections in blue and a smaller distance limited sample is shown in red. The distance limited sample includes all sources with heliocentric distances less than 3\,kpc with luminosities greater than 1000\,\lsun\ and should therefore be relatively unaffected by the Malmquist bias that can lead to false correlations. The dashed lines show the bisector least-squared fits to the data and clearly shows that the water maser luminosity is correlated with the bolometric luminosity for the whole and distance limited samples, and that the fits are in good agreement with each other. As an additional check we calculate the partial Spearman correlation coefficient using the square of the distance as an independent parameter; a value of 0.44 would suggest the correlation is real. From the linear bisector least-squares fit to all of the plotted data we obtain the following relationship:

\begin{equation}
L_{\rm{H_2O}}=a(L_{\rm{bol}})^b,
\end{equation}

\noindent where $L_{\rm{bol}}$ is the RMS bolometric luminosity and $a=(7.1\pm0.3\times10^{-12}$ and $b=(1.47\pm0.76)$. We conclude from the correlation of the cloud and water-maser velocities and the bolometric and maser luminosity that there is a strong dynamical relationship between the embedded young massive star and the H$_2$O maser as has been previously suggested from similar studies of star forming regions (e.g., \citealt{churchwell1990,anglada1996}).

\subsection{Ammonia (3,3) maser?}

\begin{figure}
\begin{center}
\includegraphics[width=0.49\textwidth, trim= 30 0 0 0]{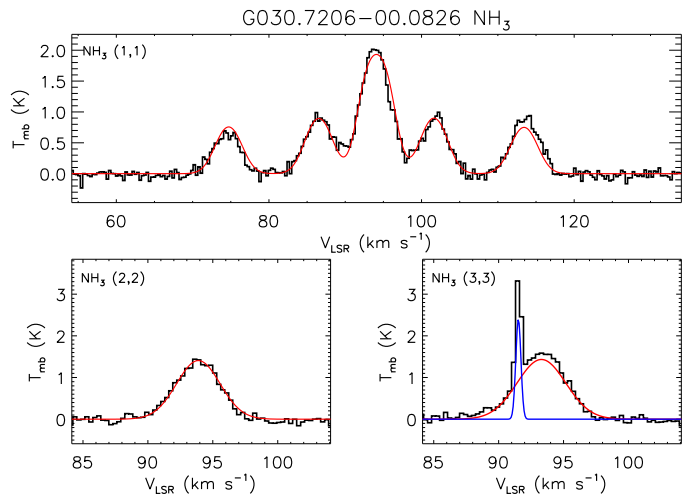}

\caption{\label{fig:nh3_maser} The NH$_3$ (1,1), (2,2) and (3,3) spectra detected towards G030.7206$-$00.0826. The NH$_3$ (3,3) emission is clearly made up from two components, a broad component thermal in nature and a stronger thin component possibly indicating the presence of an ammonia maser; the fits to these two components are coloured red and blue respectively.} 

\end{center}
\end{figure}

We detected thermal NH$_3$ (3,3) emission towards 290 sources which corresponds to approximately 50\,per\,cent of the sample observed. In all but one of these cases we found that the observed emission could be approximated by a single Gaussian profile. However, the NH$_3$ (3,3) emission detected towards the MSX source G030.7206$-$00.0826 consists of two distinct components, a broad component associated with the thermal emission from the embedded source and a second narrow-lined component. The ammonia spectra for this source are presented in Fig.\,\ref{fig:nh3_maser}.

We have fitted the NH$_3$ (3,3) emission with two Gaussian profiles to account for the broad and narrow-lined component. The fitted linewidth of the narrow component is 1.1\,\kms\ and it has a peak flux density of $\sim$5\,Jy, which makes it the narrowest NH$_3$ (3,3) line detected in the sample. Since this narrow-lined component does not have a corresponding feature seen at the same velocity in the NH$_3$ (1,1) and (2,2) emission profiles it is unlikely to be thermal in nature. Given its unusually narrow line width this is likely to be a maser transition. It is not possible to confirm the nature of this emission feature with the current data and we await further observations.

Another suspected NH$_3$ maser (G23.33$-$0.30) has recently been reported by \citet{walsh2011} from the {\bf{H$_2$O}} Southern Galactic {\bf{P}}lane {\bf{S}}urvey (HOPS) 12-mm survey (\citealt{walsh2008}).  G23.33$-$0.30 has a flux density of 9.7\,Jy, which is comparable with the flux density of the narrow-lined emission feature reported here. \citet{walsh2011} note that although the NH$_3$ (3,3) transition has been detected as a maser in a number of star forming regions (e.g., DR 21 (OH) --- \citealt{mangum1994}; W51 --- \citealt{zhang1995}; G5.89$-$0.39 --- \citealt{hunter2008}) it has never been stronger than 0.5\,Jy and thus, if these two masers can be verified they would be the strongest NH$_3$ (3,3) masers reported to date. 

G030.7206$-$00.0826 is associated with a bright unresolved radio source ($\sim$0.5\,Jy at 5\,GHz; \citealt{white2005}) and has been classified as an H{\sc ii} region with a bolometric luminosity of $\sim 4 \times 10^4$\,\lsun. \citet{walsh2011} found G23.33$-$0.30 to be positionally coincident with an IRDC and a cold dust continuum source and thus it therefore seems likely that these maser candidates arise in star forming regions. However, since only two NH$_3$ (3,3) maser candidates have been reported from the 100 square degrees surveyed by HOPS and from the GBT observations of $\sim$600 high-mass star forming regions reported here, we would conclude that bright masers of this type are relatively rare.

\section{Summary and conclusions}
\label{sect:summary}

We have used the 100-m Green Bank Telescope to conduct a programme of 22-24\,GHz spectral line observations towards 597 massive young stellar objects (MYSOs) and compact and ultra-compact H{\sc ii} regions identified by the Red MSX Source (RMS) survey. We targeted the 22\,GHz water (H$_2$O) maser transition and thermal ammonia (NH$_3$) (1,1), (2,2) and (3,3) inversion transitions.  These observations have an angular resolution of $\sim$30\arcsec\ and typical 4$\sigma$ sensitivity of $T_{\rm{mb}}= 0.2$\,K per 0.32-km\,s$^{-1}$ channel.

We detect water maser emission towards 308 RMS sources with an overall detection rate of $\sim$50\,per\,cent. NH$_3$ emission is detected towards 479 sources ($\sim$80\,per\,cent of the sample) with the NH$_3$ (1,1) hyperfine structure being clearly detected towards $\sim$400 sources, and thus allowing the kinetic temperature, optical depth and column density to be determined for a large number of massive star-forming regions. The average kinetic temperature, FWHM line width and column density for the sample are approximately 22\,K, 2\,\kms\ and $2\times 10^{15}$\,cm$^{-2}$, respectively. 

We find the detection rates for both the H$_2$O masers and NH$_3$ emission is similar for both MYSO and H{\sc ii} region sub-samples. Combining these results with a large database of complementary observations we analyse these parameters as a function of luminosity and evolutionary phase, and correlate these physical conditions with H$_2$O maser emission. Our main findings are as follows:

\begin{enumerate}

	\item There is no significant difference in the H$_2$O maser detection rate for H{\sc ii} regions and MYSOs which would suggest that the conditions required to produce maser emission are equally likely in both phases. Comparing the detection rates as a function of luminosity we find the H$_2$O detection rate has a positive dependence on the source luminosity, with the detection rate increasing with increasing luminosity.

    \item We find that the NH$_3$ (1,1) line width and kinetic temperature are correlated with luminosity. It is clear from these correlations that the embedded sources are having measurable effects on their surrounding molecular environments. Finding no underlying dependence of these parameters on the evolutionary phase of the embedded sources, we conclude that the observed trends in the derived parameters are related to the energy output of the central source and/or the linewidth-clump mass relationship. 

    \item The velocities of the peak H$_2$O masers and the NH$_3$ emission for each relevant source are in excellent agreement with each other with a correlation coefficient of 0.92, which would strongly suggest an association between the dense gas and the maser emission. Moreover, we find the bolometric luminosity of the embedded source and the isotropic luminosity of the H$_2$O maser are also correlated. We conclude that there is a strong dynamical relationship between the embedded young massive star and the H$_2$O maser.
     
    \item We find excellent agreement between velocities of sources using $^{13}$CO observations and those obtained from NH$_3$ transitions. We find only three sources where the velocity assigned using the CO data is incorrect.

\end{enumerate}

These observations are a first step in examining the global characteristics of this Galaxy-wide sample of massive young stars and will form the cornerstone for more detailed studies of well selected sub-samples.

\section*{Acknowledgments}

The authors would like to thank the Director and staff of the GBT for their assistance during the preparation of these observations. JSU is supported by a CSIRO OCE postdoctoral grant. LKM is supported by an STFC post-doctoral grant (ST/GO01847/1). The National Radio Astronomy Observatory is a facility of the National Science Foundation operated under cooperative agreement by Associated Universities, Inc. This paper made use of information from the Red MSX Source survey database at www.ast.leeds.ac.uk/RMS which was constructed with support from the Science and Technology Facilities Council of the UK.

\bibliography{scuba2}

\begin{thebibliography}{}

\bibitem[\protect\citeauthoryear{{Anglada}, {Estalella}, {Pastor}, {Rodriguez}
  \& {Haschick}}{{Anglada} et~al.}{1996}]{anglada1996}
{Anglada} G.,  {Estalella} R.,  {Pastor} J.,  {Rodriguez} L.~F.,    {Haschick}
  A.~D.,  1996, \apj, 463, 205

\bibitem[\protect\citeauthoryear{{Benjamin} \& et al.}{{Benjamin} \&
  et~al.}{2003}]{benjamin2003}
{Benjamin} R.~A.,  et al. 2003, \pasp, 115, 953

\bibitem[\protect\citeauthoryear{{Bergin} \& {Langer}}{{Bergin} \&
  {Langer}}{1997}]{bergin1997}
{Bergin} E.~A.,  {Langer} W.~D.,  1997, \apj, 486, 316

\bibitem[\protect\citeauthoryear{{Brand} \& {Blitz}}{{Brand} \&
  {Blitz}}{1993}]{brand1993}
{Brand} J.,  {Blitz} L.,  1993, \aap, 275, 67

\bibitem[\protect\citeauthoryear{{Breen}, {Caswell}, {Ellingsen} \&
  {Phillips}}{{Breen} et~al.}{2010}]{breen2010}
{Breen} S.~L.,  {Caswell} J.~L.,  {Ellingsen} S.~P.,    {Phillips} C.~J.,
  2010, \mnras, 406, 1487

\bibitem[\protect\citeauthoryear{{Bunn}, {Hoare} \& {Drew}}{{Bunn}
  et~al.}{1995}]{bunn1995}
{Bunn} J.~C.,  {Hoare} M.~G.,    {Drew} J.~E.,  1995, \mnras, 272, 346

\bibitem[\protect\citeauthoryear{{Caswell} \& {Breen}}{{Caswell} \&
  {Breen}}{2010}]{caswell2010a}
{Caswell} J.~L.,  {Breen} S.~L.,  2010, \mnras, 407, 2599

\bibitem[\protect\citeauthoryear{{Churchwell}, {Walmsley} \&
  {Cesaroni}}{{Churchwell} et~al.}{1990}]{churchwell1990}
{Churchwell} E.,  {Walmsley} C.~M.,    {Cesaroni} R.,  1990, \aaps, 83, 119

\bibitem[\protect\citeauthoryear{{Clarke}, {Lumsden}, {Oudmaijer}, {Busfield},
  {Hoare}, {Moore}, {Sheret} \& {Urquhart}}{{Clarke} et~al.}{2006}]{clarke2006}
{Clarke} A.~J.,  {Lumsden} S.~L.,  {Oudmaijer} R.~D.,  {Busfield} A.~L.,
  {Hoare} M.~G.,  {Moore} T.~J.~T.,  {Sheret} T.~L.,    {Urquhart} J.~S.,
  2006, \aap, 457, 183

\bibitem[\protect\citeauthoryear{{Claussen}, {Wilking}, {Benson}, {Wootten},
  {Myers} \& {Terebey}}{{Claussen} et~al.}{1996}]{claussen1996}
{Claussen} M.~J.,  {Wilking} B.~A.,  {Benson} P.~J.,  {Wootten} A.,  {Myers}
  P.~C.,    {Terebey} S.,  1996, \apjs, 106, 111

\bibitem[\protect\citeauthoryear{{Codella}, {Lorenzani}, {Gallego}, {Cesaroni}
  \& {Moscadelli}}{{Codella} et~al.}{2004}]{codella2004}
{Codella} C.,  {Lorenzani} A.,  {Gallego} A.~T.,  {Cesaroni} R.,
  {Moscadelli} L.,  2004, \aap, 417, 615

\bibitem[\protect\citeauthoryear{{Davies}, {Hoare}, {Lumsden}, {Hosokawa},
  {Oudmaijer}, {Urquhart}, {Mottram} \& {Stead}}{{Davies}
  et~al.}{2011}]{davies2011}
{Davies} B.,  {Hoare} M.~G.,  {Lumsden} S.~L.,  {Hosokawa} T.,  {Oudmaijer}
  R.~D.,  {Urquhart} J.~S.,  {Mottram} J.~C.,    {Stead} J.,  2011, ArXiv
  e-prints

\bibitem[\protect\citeauthoryear{{De Buizer}, {Watson}, {Radomski}, {Pi{\~n}a}
  \& {Telesco}}{{De Buizer} et~al.}{2002}]{de-buizer2002}
{De Buizer} J.~M.,  {Watson} A.~M.,  {Radomski} J.~T.,  {Pi{\~n}a} R.~K.,
  {Telesco} C.~M.,  2002, \apjl, 564, L101

\bibitem[\protect\citeauthoryear{{Evans} II}{{Evans}}{1999}]{evans1999}
{Evans} II N.~J.,  1999, \araa, 37, 311

\bibitem[\protect\citeauthoryear{{Forster} \& {Caswell}}{{Forster} \&
  {Caswell}}{1999}]{forster1999}
{Forster} J.~R.,  {Caswell} J.~L.,  1999, \aaps, 137, 43

\bibitem[\protect\citeauthoryear{{Ho} \& {Townes}}{{Ho} \&
  {Townes}}{1983}]{ho1983}
{Ho} P.~T.~P.,  {Townes} C.~H.,  1983, \araa, 21, 239

\bibitem[\protect\citeauthoryear{{Hoare}, {Drew}, {Muxlow} \& {Davis}}{{Hoare}
  et~al.}{1994}]{hoare1994}
{Hoare} M.~G.,  {Drew} J.~E.,  {Muxlow} T.~B.,    {Davis} R.~J.,  1994, \apjl,
  421, L51

\bibitem[\protect\citeauthoryear{{Hoare}, {Lumsden}, {Oudmaijer}, {Urquhart},
  {Busfield}, {Sheret}, {Clarke}, {Moore}, {Allsopp}, {Burton}, {Purcell},
  {Jiang} \& {Wang}}{{Hoare} et~al.}{2005}]{hoare2005}
{Hoare} M.~G.,  {Lumsden} S.~L.,  {Oudmaijer} R.~D.,  {Urquhart} J.~S.,
  {Busfield} A.~L.,  {Sheret} T.~L.,  {Clarke} A.~J.,  {Moore} T.~J.~T.,
  {Allsopp} J.,  {Burton} M.~G.,  {Purcell} C.~R.,  {Jiang} Z.,    {Wang} M.,
  2005, in {Cesaroni} R.,  {Felli} M.,  {Churchwell} E.,   {Walmsley} M.,  eds,
  Massive Star Birth: A Crossroads of Astrophysics Vol.~227 of IAU Symposium,
  {The RMS survey: Massive young stars throughout the galaxy}.
pp 370--375

\bibitem[\protect\citeauthoryear{{Hosokawa} \& {Omukai}}{{Hosokawa} \&
  {Omukai}}{2009}]{hosokawa2009}
{Hosokawa} T.,  {Omukai} K.,  2009, \apj, 691, 823

\bibitem[\protect\citeauthoryear{{Hosokawa}, {Yorke} \& {Omukai}}{{Hosokawa}
  et~al.}{2010}]{hosokawa2010}
{Hosokawa} T.,  {Yorke} H.~W.,    {Omukai} K.,  2010, \apj, 721, 478

\bibitem[\protect\citeauthoryear{{Hunter}, {Brogan}, {Indebetouw} \&
  {Cyganowski}}{{Hunter} et~al.}{2008}]{hunter2008}
{Hunter} T.~R.,  {Brogan} C.~L.,  {Indebetouw} R.,    {Cyganowski} C.~J.,
  2008, \apj, 680, 1271

\bibitem[\protect\citeauthoryear{{Kurtz} \& {Hofner}}{{Kurtz} \&
  {Hofner}}{2005}]{kurtz2005}
{Kurtz} S.,  {Hofner} P.,  2005, \aj, 130, 711

\bibitem[\protect\citeauthoryear{{Lada}}{{Lada}}{1985}]{lada1985}
{Lada} C.~J.,  1985, \araa, 23, 267

\bibitem[\protect\citeauthoryear{{Larson}}{{Larson}}{1981}]{larson1981}
{Larson} R.~B.,  1981, \mnras, 194, 809

\bibitem[\protect\citeauthoryear{{Li}, {Goldsmith} \& {Menten}}{{Li}
  et~al.}{2003}]{li2003}
{Li} D.,  {Goldsmith} P.~F.,    {Menten} K.,  2003, \apj, 587, 262

\bibitem[\protect\citeauthoryear{{Longmore}, {Burton}, {Barnes}, {Wong},
  {Purcell} \& {Ott}}{{Longmore} et~al.}{2007}]{longmore2007}
{Longmore} S.~N.,  {Burton} M.~G.,  {Barnes} P.~J.,  {Wong} T.,  {Purcell}
  C.~R.,    {Ott} J.,  2007, \mnras, 379, 535

\bibitem[\protect\citeauthoryear{{Mangum} \& {Wootten}}{{Mangum} \&
  {Wootten}}{1994}]{mangum1994}
{Mangum} J.~G.,  {Wootten} A.,  1994, \apjl, 428, L33

\bibitem[\protect\citeauthoryear{{Mangum}, {Wootten} \& {Mundy}}{{Mangum}
  et~al.}{1992}]{mangum1992}
{Mangum} J.~G.,  {Wootten} A.,    {Mundy} L.~G.,  1992, \apj, 388, 467

\bibitem[\protect\citeauthoryear{{McKee} \& {Tan}}{{McKee} \&
  {Tan}}{2003}]{mckee2003}
{McKee} C.~F.,  {Tan} J.~C.,  2003, \apj, 585, 850

\bibitem[\protect\citeauthoryear{{Mottram}, {Hoare}, {Davies}, {Lumsden},
  {Oudmaijer}, {Urquhart}, {Moore}, {Cooper} \& {Stead}}{{Mottram}
  et~al.}{2011}]{mottram2011b}
{Mottram} J.~C.,  {Hoare} M.~G.,  {Davies} B.,  {Lumsden} S.~L.,  {Oudmaijer}
  R.~D.,  {Urquhart} J.~S.,  {Moore} T.~J.~T.,  {Cooper} H.~D.~B.,    {Stead}
  J.~J.,  2011, \apjl, 730, L33+

\bibitem[\protect\citeauthoryear{{Mottram}, {Hoare}, {Lumsden}, {Oudmaijer},
  {Urquhart}, {Sheret}, {Clarke} \& {Allsopp}}{{Mottram}
  et~al.}{2007}]{mottram2007}
{Mottram} J.~C.,  {Hoare} M.~G.,  {Lumsden} S.~L.,  {Oudmaijer} R.~D.,
  {Urquhart} J.~S.,  {Sheret} T.~L.,  {Clarke} A.~J.,    {Allsopp} J.,  2007,
  \aap, 476, 1019

\bibitem[\protect\citeauthoryear{{Mottram}, {Hoare}, {Urquhart}, {Lumsden},
  {Oudmaijer}, {Robitaille}, {Moore}, {Davies} \& {Stead}}{{Mottram}
  et~al.}{2011}]{mottram2011a}
{Mottram} J.~C.,  {Hoare} M.~G.,  {Urquhart} J.~S.,  {Lumsden} S.~L.,
  {Oudmaijer} R.~D.,  {Robitaille} T.~P.,  {Moore} T.~J.~T.,  {Davies} B.,
  {Stead} J.,  2011, \aap, 525, A149+

\bibitem[\protect\citeauthoryear{{Mottram}, {Urquhart}, {Hoare}, {Lumsden} \&
  {Oudmaijer}}{{Mottram} et~al.}{2006}]{mottram2006}
{Mottram} J.~C.,  {Urquhart} J.~S.,  {Hoare} M.~G.,  {Lumsden} S.~L.,
  {Oudmaijer} R.~D.,  2006, ArXiv Astrophysics e-prints

\bibitem[\protect\citeauthoryear{{Pillai}, {Wyrowski}, {Carey} \&
  {Menten}}{{Pillai} et~al.}{2006}]{pillai2006}
{Pillai} T.,  {Wyrowski} F.,  {Carey} S.~J.,    {Menten} K.~M.,  2006, \aap,
  450, 569

\bibitem[\protect\citeauthoryear{{Rosolowsky}, {Pineda}, {Foster}, {Borkin},
  {Kauffmann}, {Caselli}, {Myers} \& {Goodman}}{{Rosolowsky}
  et~al.}{2008}]{Rosolowsky2008}
{Rosolowsky} E.~W.,  {Pineda} J.~E.,  {Foster} J.~B.,  {Borkin} M.~A.,
  {Kauffmann} J.,  {Caselli} P.,  {Myers} P.~C.,    {Goodman} A.~A.,  2008,
  \apjs, 175, 509

\bibitem[\protect\citeauthoryear{{Rydbeck}, {Sume}, {Hjalmarson}, {Ellder},
  {Ronnang} \& {Kollberg}}{{Rydbeck} et~al.}{1977}]{rydbeck1977}
{Rydbeck} O.~E.~H.,  {Sume} A.,  {Hjalmarson} A.,  {Ellder} J.,  {Ronnang}
  B.~O.,    {Kollberg} E.,  1977, \apjl, 215, L35

\bibitem[\protect\citeauthoryear{{Sridharan}, {Beuther}, {Saito}, {Wyrowski} \&
  {Schilke}}{{Sridharan} et~al.}{2005}]{sridharan2005}
{Sridharan} T.~K.,  {Beuther} H.,  {Saito} M.,  {Wyrowski} F.,    {Schilke} P.,
   2005, \apjl, 634, L57

\bibitem[\protect\citeauthoryear{{Stahler} \& {Palla}}{{Stahler} \&
  {Palla}}{2005}]{stahler2005}
{Stahler} S.~W.,  {Palla} F.,  2005, {The Formation of Stars}

\bibitem[\protect\citeauthoryear{{Swift}, {Welch} \& {Di Francesco}}{{Swift}
  et~al.}{2005}]{swift2005}
{Swift} J.~J.,  {Welch} W.~J.,    {Di Francesco} J.,  2005, \apj, 620, 823

\bibitem[\protect\citeauthoryear{{Urquhart}, {Busfield}, {Hoare}, {Lumsden},
  {Clarke}, {Moore}, {Mottram} \& {Oudmaijer}}{{Urquhart}
  et~al.}{2007}]{urquhart_radio_south}
{Urquhart} J.~S.,  {Busfield} A.~L.,  {Hoare} M.~G.,  {Lumsden} S.~L.,
  {Clarke} A.~J.,  {Moore} T.~J.~T.,  {Mottram} J.~C.,    {Oudmaijer} R.~D.,
  2007, \aap, 461, 11

\bibitem[\protect\citeauthoryear{{Urquhart}, {Busfield}, {Hoare}, {Lumsden},
  {Oudmaijer}, {Moore}, {Gibb}, {Purcell}, {Burton} \& {Marechal}}{{Urquhart}
  et~al.}{2007}]{urquhart_13co_south}
{Urquhart} J.~S.,  {Busfield} A.~L.,  {Hoare} M.~G.,  {Lumsden} S.~L.,
  {Oudmaijer} R.~D.,  {Moore} T.~J.~T.,  {Gibb} A.~G.,  {Purcell} C.~R.,
  {Burton} M.~G.,    {Marechal} L.~J.~L.,  2007, \aap, 474, 891

\bibitem[\protect\citeauthoryear{{Urquhart}, {Busfield}, {Hoare}, {Lumsden},
  {Oudmaijer}, {Moore}, {Gibb}, {Purcell}, {Burton}, {Mar{\'e}chal}, {Jiang} \&
  {Wang}}{{Urquhart} et~al.}{2008}]{urquhart_13co_north}
{Urquhart} J.~S.,  {Busfield} A.~L.,  {Hoare} M.~G.,  {Lumsden} S.~L.,
  {Oudmaijer} R.~D.,  {Moore} T.~J.~T.,  {Gibb} A.~G.,  {Purcell} C.~R.,
  {Burton} M.~G.,  {Mar{\'e}chal} L.~J.~L.,  {Jiang} Z.,    {Wang} M.,  2008,
  \aap, 487, 253

\bibitem[\protect\citeauthoryear{{Urquhart}, {Hoare}, {Lumsden}, {Oudmaijer} \&
  {Moore}}{{Urquhart} et~al.}{2008}]{urquhart2007c}
{Urquhart} J.~S.,  {Hoare} M.~G.,  {Lumsden} S.~L.,  {Oudmaijer} R.~D.,
  {Moore} T.~J.~T.,  2008, in {Beuther} H.,  {Linz} H.,   {Henning} T.,  eds,
  Massive Star Formation: Observations Confront Theory Vol.~387 of Astronomical
  Society of the Pacific Conference Series, {The RMS Survey: A Galaxy-wide
  Sample of Massive Young Stellar Objects}.
pp 381--+

\bibitem[\protect\citeauthoryear{{Urquhart}, {Hoare}, {Lumsden}, {Oudmaijer},
  {Moore}, {Brook}, {Mottram}, {Davies} \& {Stead}}{{Urquhart}
  et~al.}{2009}]{urquhart2009_h2o}
{Urquhart} J.~S.,  {Hoare} M.~G.,  {Lumsden} S.~L.,  {Oudmaijer} R.~D.,
  {Moore} T.~J.~T.,  {Brook} P.~R.,  {Mottram} J.~C.,  {Davies} B.,    {Stead}
  J.~J.,  2009, \aap, 507, 795

\bibitem[\protect\citeauthoryear{{Urquhart}, {Hoare}, {Purcell}, {Lumsden},
  {Oudmaijer}, {Moore}, {Busfield}, {Mottram} \& {Davies}}{{Urquhart}
  et~al.}{2009}]{urquhart_radio_north}
{Urquhart} J.~S.,  {Hoare} M.~G.,  {Purcell} C.~R.,  {Lumsden} S.~L.,
  {Oudmaijer} R.~D.,  {Moore} T.~J.~T.,  {Busfield} A.~L.,  {Mottram} J.~C.,
  {Davies} B.,  2009, \aap, 501, 539

\bibitem[\protect\citeauthoryear{{Urquhart}, {Moore}, {Hoare}, {Lumsden},
  {Oudmaijer}, {Rathborne}, {Mottram}, {Davies} \& {Stead}}{{Urquhart}
  et~al.}{2011}]{urquhart2011}
{Urquhart} J.~S.,  {Moore} T.~J.~T.,  {Hoare} M.~G.,  {Lumsden} S.~L.,
  {Oudmaijer} R.~D.,  {Rathborne} J.~M.,  {Mottram} J.~C.,  {Davies} B.,
  {Stead} J.~J.,  2011, \mnras, 410, 1237

\bibitem[\protect\citeauthoryear{{Walmsley} \& {Ungerechts}}{{Walmsley} \&
  {Ungerechts}}{1983}]{walmsley1983}
{Walmsley} C.~M.,  {Ungerechts} H.,  1983, \aap, 122, 164

\bibitem[\protect\citeauthoryear{{Walsh}, {Breen}, {Britton}, {Brooks},
  {Burton}, {Cunningham}, {Green}, {Harvey-Smith}, {Hindson}, {Hoare},
  {Indermuehle}, {Jones}, {Lo}, {Longmore}, {Lowe}, {Phillips}, {Purcell} \&
  five~other authors}{{Walsh} et~al.}{2011}]{walsh2011}
{Walsh} A.~J.,  {Breen} S.~L.,  {Britton} T.,  {Brooks} K.~J.,  {Burton} M.~G.,
   {Cunningham} M.~R.,  {Green} J.~A.,  {Harvey-Smith} L.,  {Hindson} L.,
  {Hoare} M.~G.,  {Indermuehle} B.,  {Jones} P.~A.,  {Lo} N.,  {Longmore}
  S.~N.,  {Lowe} V.,  {Phillips} C.~J.,  {Purcell} C.~R.,    five~other authors
  2011, ArXiv e-prints

\bibitem[\protect\citeauthoryear{{Walsh}, {Lo}, {Burton}, {White}, {Purcell},
  {Longmore}, {Phillips} \& {Brooks}}{{Walsh} et~al.}{2008}]{walsh2008}
{Walsh} A.~J.,  {Lo} N.,  {Burton} M.~G.,  {White} G.~L.,  {Purcell} C.~R.,
  {Longmore} S.~N.,  {Phillips} C.~J.,    {Brooks} K.~J.,  2008, Publications
  of the Astronomical Society of Australia, 25, 105

\bibitem[\protect\citeauthoryear{{White}, {Becker} \& {Helfand}}{{White}
  et~al.}{2005}]{white2005}
{White} R.~L.,  {Becker} R.~H.,    {Helfand} D.~J.,  2005, \aj, 130, 586

\bibitem[\protect\citeauthoryear{{Yorke} \& {Bodenheimer}}{{Yorke} \&
  {Bodenheimer}}{2008}]{yorke2008}
{Yorke} H.~W.,  {Bodenheimer} P.,  2008, in {H.~Beuther, H.~Linz, \&
  T.~Henning} ed., Massive Star Formation: Observations Confront Theory
  Vol.~387 of Astronomical Society of the Pacific Conference Series,
  {Theoretical Developments in Understanding Massive Star Formation}.
pp 189--+

\bibitem[\protect\citeauthoryear{{Zhang}, {Ho}, {Wright} \& {Wilner}}{{Zhang}
  et~al.}{1995}]{zhang1995}
{Zhang} Q.,  {Ho} P.~T.~P.,  {Wright} M.~C.~H.,    {Wilner} D.~J.,  1995,
  \apjl, 451, L71+

\end{thebibliography}

\bibliographystyle{mn2e}


\clearpage
\appendix

\section{Complex ammonia spectra}
\label{app:complex_spectra}

\begin{figure*}
\begin{center}
\includegraphics[width=0.4\textwidth, trim= 0 0 0 0]{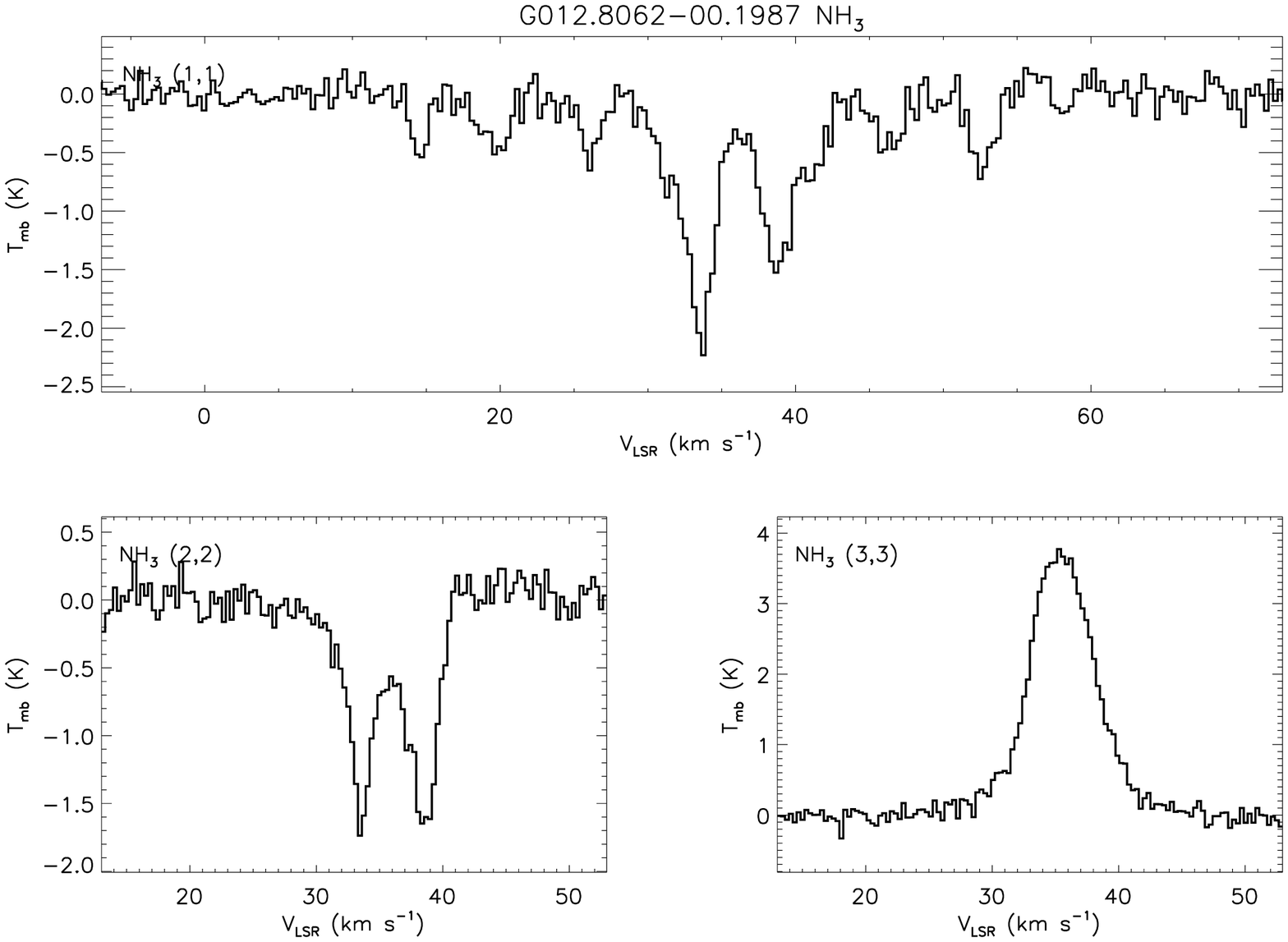}
\includegraphics[width=0.4\textwidth, trim= 0 0 0 0]{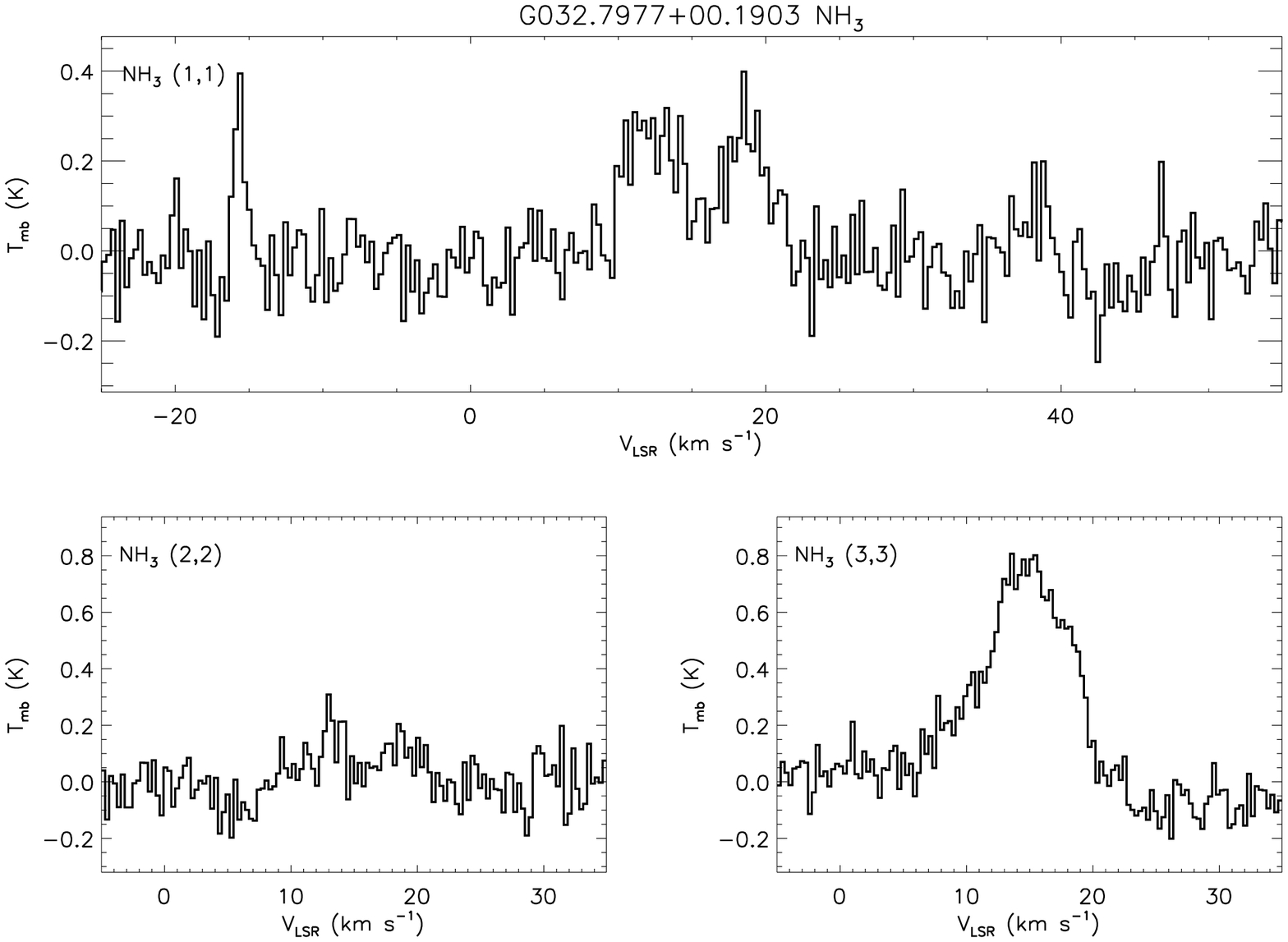}\\
\includegraphics[width=0.4\textwidth, trim= 0 0 0 0]{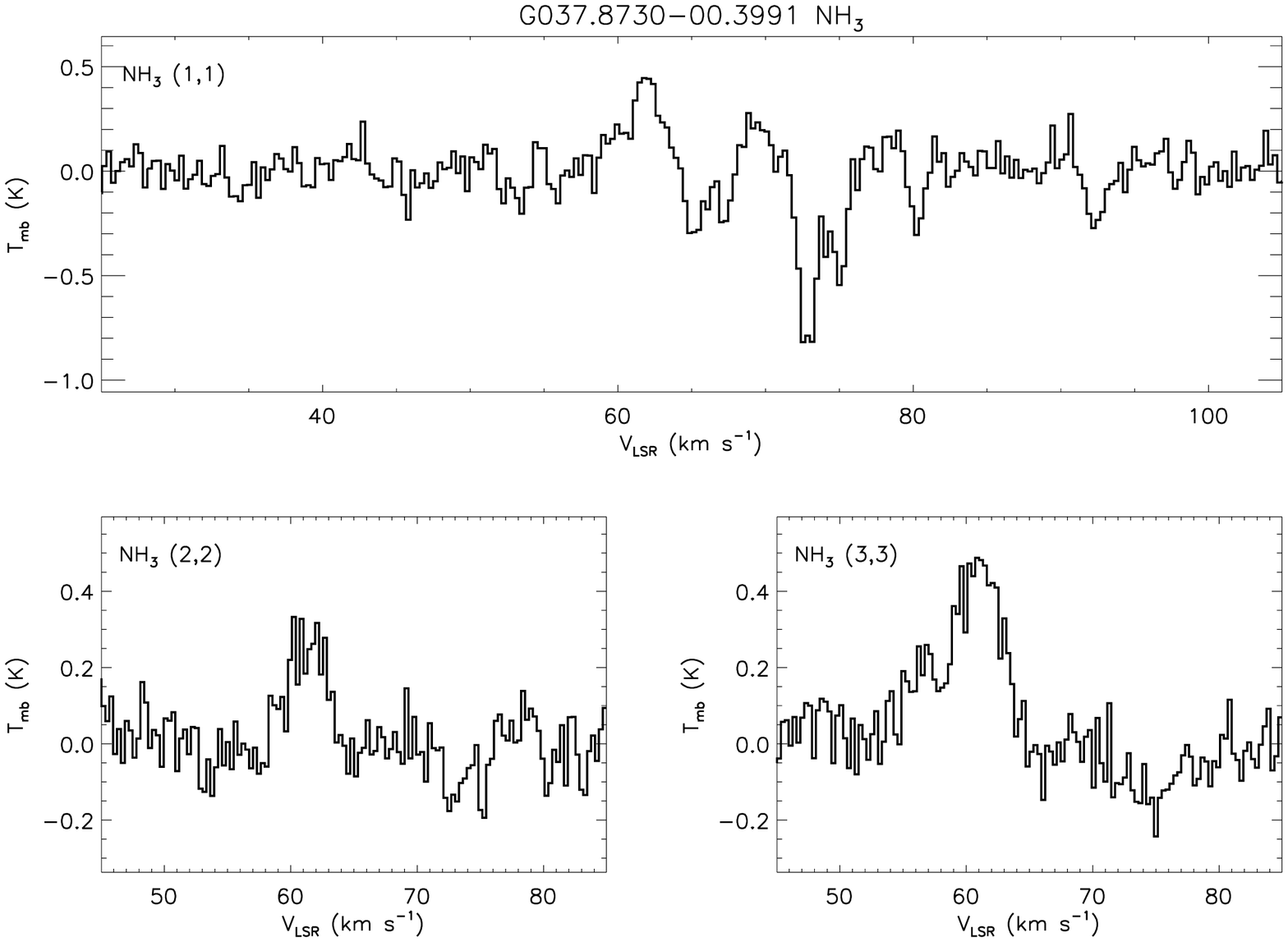}
\includegraphics[width=0.4\textwidth, trim= 0 0 0 0]{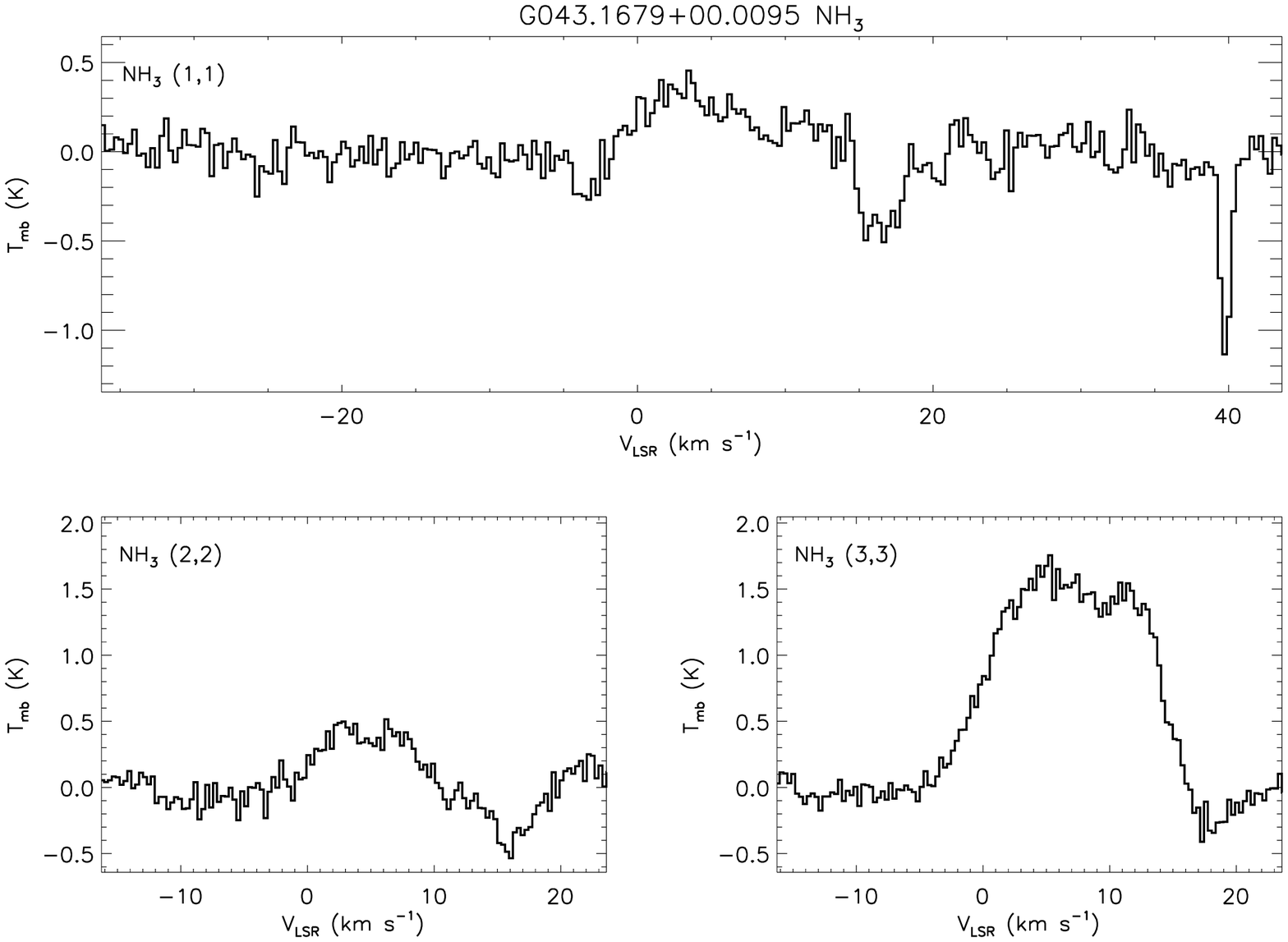}\\
\includegraphics[width=0.4\textwidth, trim= 0 0 0 0]{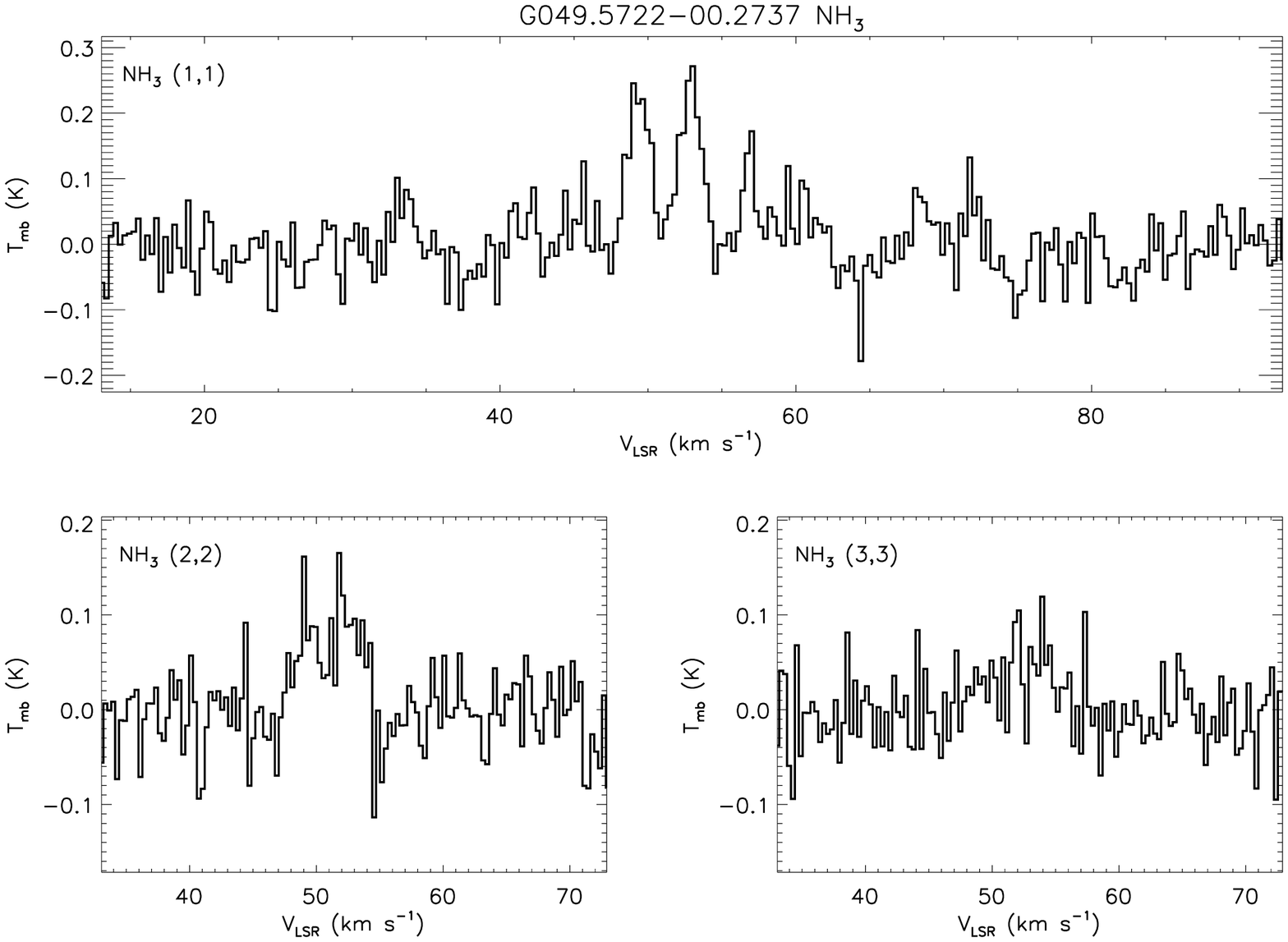}
\includegraphics[width=0.4\textwidth, trim= 0 0 0 0]{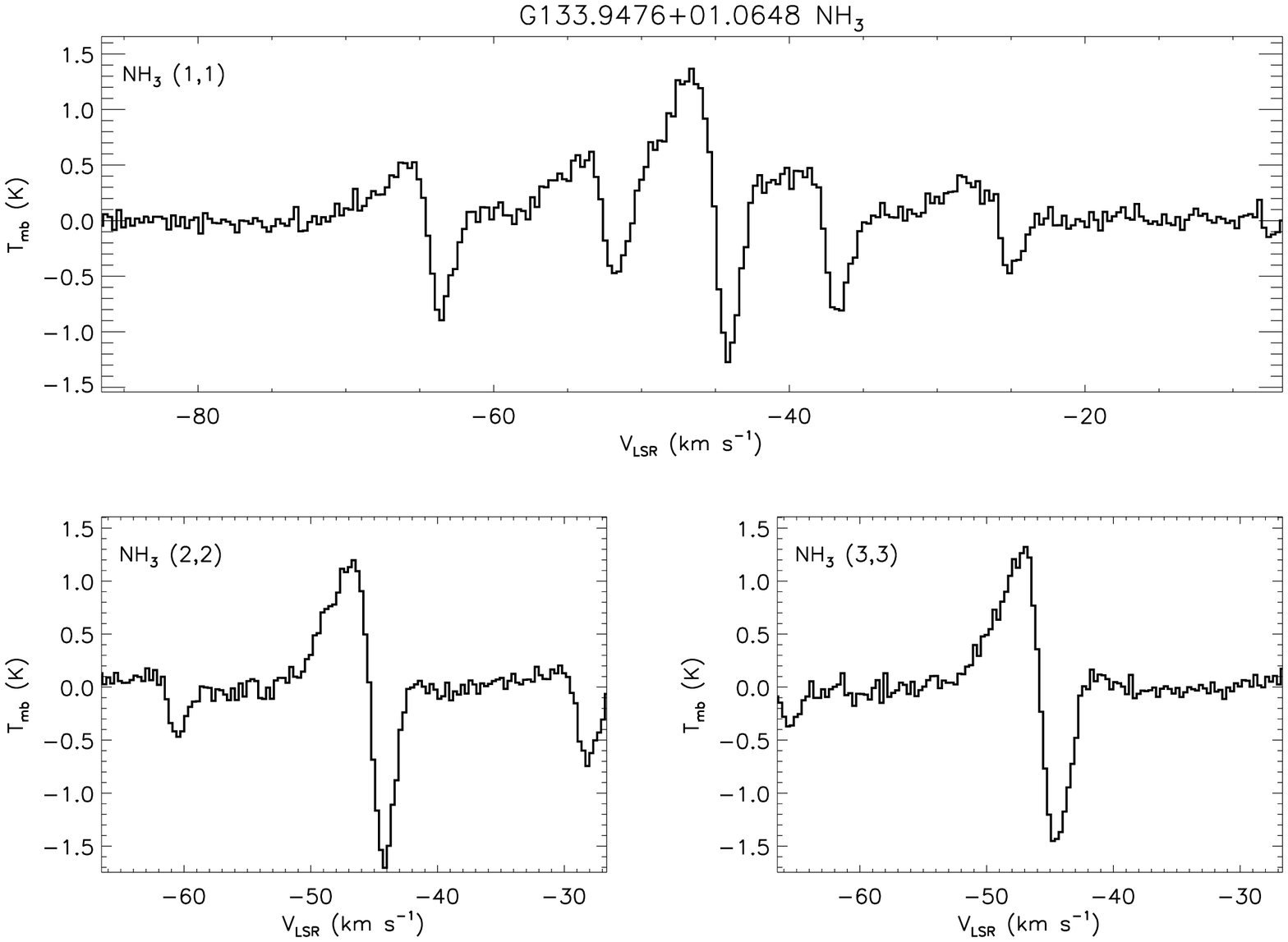}

\caption{\label{fig:nh3_spectra_complex} Complex spectra obtain towards six RMS sources that show evidence of both emission and absorption.} 

\end{center}
\end{figure*}

\end{document}